\documentclass[12pt]{article}

\usepackage{fancyhdr}

\renewcommand{\sectionmark}[1]{}
\renewcommand{\subsectionmark}[1]{}

\usepackage[utf8]{inputenc}
\usepackage{geometry}
\usepackage{natbib}
\usepackage{amsmath}
\usepackage{amssymb}
\usepackage{hyperref}
\usepackage{bookt abs}
\usepackage[table]{xcolor}
\usepackage{bm}
\usepackage{pdflscape}
\usepackage{tabu}
\usepackage{subcaption}
\usepackage{setspace}
\usepackage{needspace}
\hypersetup{colorlinks=true,linkcolor=blue,filecolor=blue,citecolor=blue}

\usepackage{tikz}
\usepackage[labelfont=bf]{caption}
\usepackage{relsize}
\usepackage{xifthen}

\usepackage{titlesec}
\titlespacing\section{0pt}{\parskip}{\parskip}

\allowdisplaybreaks

\usepackage{multibib}
\newcites{S}{Web Appendix References}

\newtheorem{theorem}{Theorem}
\newtheorem{assumption}{Assumption}

\newtheorem{intassumption}{Assumption}
\numberwithin{intassumption}{assumption}

\providecommand{\keywords}[1]
{
  \small	
  \textit{Keywords---} #1
}

\newcommand{\E}{\mathbb{E}}
\newcommand{\PP}{\mathbb{P}}
\newcommand{\bX}{\boldsymbol X}
\newcommand{\bx}{\boldsymbol x}
\newcommand{\indep}{\perp \!\!\! \perp}
\newcommand{\cp}{\overset{p}{\rightarrow}}
\newcommand{\val}{\text{val}}
\newcommand{\main}{\text{main}}
\newcommand{\eqdef}{\overset{\text{def}}{=}}
\newcommand{\tate}{\text{TATE}}
\newcommand{\ep}{\text{e.p.}}
\newcommand{\Var}{\text{Var}}
\newcommand{\Cov}{\text{Cov}}

\newcommand{\Scal}{\mathcal{S}}
\newcommand{\cv}{\text{CV}}

\usepackage{parskip}
\expandafter\def\expandafter\normalsize\expandafter{%
    \normalsize%
    \setlength\abovedisplayskip{5pt}%
    \setlength\belowdisplayskip{5pt}%
    \setlength\abovedisplayshortskip{-8pt}%
    \setlength\belowdisplayshortskip{2pt}%
}

\DeclareCaptionLabelFormat{webfig}{Web Appendix Figure #2}
\DeclareCaptionLabelFormat{webtab}{Web Appendix Table #2}

\title{\vspace{-3em} \fontsize{18pt}
{18pt}\selectfont Flexible and Efficient Estimation of Causal Effects with Error-Prone Exposures: A Control Variates Approach for Measurement Error}
\author{\fontsize{12pt}{12pt}\selectfont Keith Barnatchez$^{1,*}$, Rachel Nethery$^1$, Bryan E. Shepherd$^2$, Giovanni Parmigiani$^{1,3}$,  \\ \fontsize{12pt}{12pt}\selectfont Kevin P. Josey$^4$}
\date{\fontsize{12pt}{16pt}\selectfont $^1$Department of Biostatistics, Harvard T.H. Chan School of Public Health \\
$^2$Department of Biostatistics, Vanderbilt University Medical Center 
\\
$^3$Department of Data Science, Dana-Farber Cancer Institute \\
$^4$Department of Biostatistics and Informatics, Colorado School of Public Health
\\[0.75em]
\today
\\[0.75em]
\fontsize{10pt}{10pt}\selectfont
*\textit{Email:} keithbarnatchez@g.harvard.edu
}

\newcommand{\showfig}{1}

\begin{document}

 \maketitle
\normalsize

\begin{abstract}
Exposure measurement error is a ubiquitous but often overlooked challenge in causal inference with observational data. Existing methods accounting for exposure measurement error largely rely on restrictive parametric assumptions, while emerging data-adaptive estimation approaches allow for less restrictive assumptions but at the cost of flexibility, as they are typically tailored towards rigidly-defined statistical quantities. There remains a critical need for assumption-lean estimation methods that are both flexible and possess desirable theoretical properties across a variety of study designs. In this paper, we introduce a general framework for estimation of causal quantities in the presence of exposure measurement error, adapted from the method of control variates. Our method can be implemented in various two-phase sampling study designs, where one obtains   gold-standard exposure measurements for a small subset of the full study sample, called the validation data. The control variates framework leverages both the error-prone and error-free exposure measurements by augmenting an initial consistent estimator from the validation data with a variance reduction term formed from the full data. We show that our method inherits double-robustness properties under standard causal assumptions. Simulation studies show that our approach performs favorably compared to leading methods under various two-phase sampling schemes. We illustrate our method with observational electronic health record data on HIV outcomes from the Vanderbilt Comprehensive Care Clinic.
\end{abstract}

\doublespacing

\keywords{Causal inference, Control variates, Doubly-robust,  HIV, Measurement error,   Validation data}
\thispagestyle{empty}
\setcounter{page}{0}
\clearpage
\normalsize
\section{Introduction}
\label{sec:intro}
Measurement error poses a significant challenge in public health applications ranging from environmental health to nutritional epidemiology. 
Examples include area-level air pollution measurements extrapolated from sensors in fixed locations, self-reported dietary consumption patterns deviating considerably from true consumption patterns, and reported adherence to daily-regimen treatments like pre-exposure prophylaxis, which are often biased.
These three scenarios are just a few examples of an overarching problem in observational and experimental studies -- the exposure of interest is often difficult and/or expensive to measure correctly. In these scenarios, researchers often resort to using 
error-prone measurements as proxies of the true exposure. In the context of parametric models, it is well-understood that measurement error in an exposure of interest, if ignored, can lead to severely biased inferences (\citealt{carroll2006measurement}).

Properly addressing measurement error is particularly important
given the widespread use of electronic health record (EHR) data in biomedical research. As a motivating example, the Vanderbilt Comprehensive Care Clinic (VCCC), an outpatient facility serving individuals living with HIV, collects extensive data from patient EHRs. Using data extracted from the VCCC EHR, researchers aim to estimate the average causal effect of experiencing an AIDS-defining event (ADE) prior to a patient's first visit on subsequent five-year mortality. Chart reviews validating the EHR-derived ADE data have revealed substantial inaccuracies including false negatives of 3\% and notably higher false positives of 42\%. Analyses using the error-prone ADE measurements will result in biased and potentially clinically misleading estimates of the average causal effect of ADEs on mortality, highlighting the necessity for analytical strategies that explicitly account for measurement error. Our goal is to leverage carefully validated measurements from a subset of patient records to produce consistent and efficient estimates of this causal effect.

While a rich literature exists on statistical methods for analyzing error-prone data (\citealt{carroll2006measurement}), measurement error methods have only recently been situated within a formal causal inference framework. In the causal inference context, methods are generally designed specifically to address measurement error in either the exposure variable, the outcome variable, or the confounding variables (\citealt{valeri2021measurement}). Much of the focus has been on parametric, propensity score-based causal analyses with binary exposures (\citealt{braun2017propensity}), though recent work has extended methods to continuous (\citealt{josey2023estimating}) and categorical error-prone exposures (\citealt{wu2019causal}).
While in this work we focus on exposure measurement error and misclassification, there have been related developments in the causal inference literature for addressing differential measurement error in the outcome variable (\citealt{ackerman2021calibrating,kallus2025role}), as well as measurement error in confounding variables. In particular, measurement error in confounding variables has been extensively studied; \cite{webb2017imputation} consider multiple imputation-based methods, \cite{kyle2016simexmsm} extend the simulation extrapolation (SIMEX) method to account for measurement error in time-varying confounders in marginal structural models, and \cite{hong2017bayesian} consider Bayesian approaches to addressing confounder measurement error.
\\ \\
To date, much of the literature at the intersection of causal inference and measurement error has considered methods that rely on parametric models for the measurement error mechanism, the treatment assignment, and the outcome. This trend stems mainly from the long-established measurement error literature, which has primarily focused on bias corrections in parametric models (\citealt{wang2021identifiability}). In turn, much of the work in causal inference has ostensibly borrowed from  developments in the measurement error literature with the focus directed on inference in the context of parametric data-generating processes. While parametric modeling has played a key role in the development of causal methods, the disproportionate number of methods based on parametric modeling in measurement error applications is largely out of step with recent developments in the causal inference literature that emphasize targeting explicitly defined estimands without the reliance on assumption-heavy models.
\\ \\
In recent years, the causal inference community has made encouraging progress towards the development of robust methods that make fewer assumptions about the underlying data generating process in measurement error problems. In particular,  recent developments have been spurred by approaches that recast measurement error as a missing data problem, allowing one to implement existing tools and theory to derive improved estimators (\citealt{keogh2021measurement}). As an example of the modern robust missing data methods that could be adapted to address measurement error problems, \cite{kennedy2020efficient} proposed efficient, doubly robust methods for estimating average treatment effects under partially missing exposure information. Their method is readily adaptable to scenarios where error prone observations are treated as missing data. In a similar spirit, but in the context of partially missing outcomes, \cite{kallus2025role} developed semi-parametric efficient estimators for estimating average treatment effects that can be leveraged in scenarios where the outcome of interest is measured with error. While these methods have attractive theoretical properties, their adoption into applied research has been slow, largely due to a lack of flexibility. Since these methods target specific statistical estimands, small tweaks to the estimation problem often yield a vastly different estimator. In turn, there is a need for methods that possess the attractive properties typical of doubly-robust estimators, while accommodating numerous study designs and potential sources of bias in a general manner that facilitates their uptake.
\\ \\
Despite the recent progress in methods for causal inference with measurement error, there remains a critical need for methods that are easy to implement in a broad range of measurement error problems and study designs. In this paper, we address these needs by proposing a general framework for estimation of causal quantities in the presence of exposure measurement error through adaptations of the control variates framework developed by \cite{yang2019combining}. The control variates framework leverages a subset of validated exposure measurements to identify and correct for biases due to measurement error and to quantify the correlation between the gold-standard and error-prone effect estimates, which is used in combination with the information from the full sample to achieve an estimator with reduced variance. We show that these estimators perform competitively with existing estimators under common two-phase sampling schemes. Simulation studies show that our method performs similarly to commonly-used methods for addressing measurement error, while additionally possessing the ability to handle 
study designs for which the current leading approaches are not well-suited. Moreover, our method is straightforward to implement, only requiring small augmentations to existing popular software tools for conducting causal inference research.
\\ \\
The remainder of this paper is structured as follows. In Section \ref{probsetting}, we define the problem setting, causal estimands of interest and relevant assumptions. 
Section \ref{methods} introduces our proposed control variates method under a simplified scenario where the validation data are randomly sampled through a two-phase sampling scheme. Section \ref{simmy} presents the results of a simulation study comparing the control variates method to existing popular measurement error correction approaches. In Section \ref{application}, we assess the performance of our proposed method in real-world settings by applying it to data from the VCCC. 
Finally, Section \ref{disc} concludes with a discussion of our findings and avenues for future research.
\vspace{-0.75em}
\section{Problem Setting} 
\label{probsetting}
\vspace{-1em}
\subsection{Data and Notation}
Suppose a researcher obtains independent  samples $\bm O_i = (Y_i,A_i,A_i^*,\bX_i,S_i), \ i \in \{1,\ldots,n\}$, from a target population of interest. $A_i^*$ is an \textit{error-prone} measurement of the true binary exposure indicator $A_i$. We assume access to an \textit{internal} validation sample for which gold-standard measurements of the true exposure  $A_i$ are available. Selection into the internal validation sample is denoted by the binary variable $S_i$, and accordingly $A_i$ is  observed when $S_i=1$ and missing otherwise. $Y_i$ denotes the outcome of interest, and $\bX_i$ 
a vector of covariates, which are both initially assumed to be measured without error. Throughout, we adopt the Rubin potential outcomes framework (\citealt{rubin1974estimating}), letting $Y_i(a)$ denote the outcome subject $i$ would have experienced had they received treatment $A_i=a, \ a \in \{0,1\}$. Importantly, we define potential outcomes in terms of the true treatment value, not their error-prone measurements.
\\[1.4em]
While the availability of a validation dataset---joint observations of error-prone and error-free measurements of an exposure---may seem relatively uncommon, there are numerous instances of such data structures in applied research. Consider \textit{two-phase sampling} schemes (\citealt{carroll2006measurement}), which are often employed when measurement of key variables is difficult. In these schemes, a large dataset is initially sampled from a target population. Then for a subset of the original sample, gold-standard measurements of the difficult-to-measure variables are obtained. Moreover, this structure is often seen in studies where subjects with error-prone measurements from a main dataset can be linked to gold-standard measurements from an external source. One common example occurs within medical claims data, when investigators can link patients with error-prone medical claims data to external data sources with more detailed information. 
We provide numerous examples of studies that have made use of validation data to address measurement error in the Supplementary Materials.
\\ \\
We assume that the true and error-prone exposure measurements are related by a particular variant of a \textit{classical, differential} measurement error model. Such models 1) assume that the true exposure value is a direct cause of the error-prone measurement
and 2) allow for possible correlation between the severity of the measurement errors and the observed outcomes $Y$.  Critically, and contrary to much of the measurement error literature, we make no \textit{a priori} assumptions on the functional form of the measurement error mechanism that relates $A$ to $A^*$. As will be discussed in Section \ref{cvmethod}, our proposed approach circumvents the modeling of this process by leveraging the correlation between the true and error-prone variables.
\vspace{-1em}
\subsection{Causal Estimand}
While our proposed methodology is applicable to general functions of counterfactual means, for clarity we fix our interests in estimating the \textit{target average treatment effect} (TATE):
\[
\tau_\text{TATE} \eqdef \mathbb{E}[Y(1) - Y(0)], \] the average treatment effect of the exposure on the outcome within the population from which the main data are sampled. The distinction placed on the target population is necessary because, even if the full sample is a random sample of the target population, it may not be the case that the validation sample is.
Recall that $A_i$ is partially unobserved in the main study data. Since we wish to avoid assumptions about the structure of the measurement error, we begin by discussing the conditions needed to identify $\tau_\text{TATE}$ using only the validation data, where $A_i$ is observed.
\vspace{-0.5em}
\begin{assumption}[SUTVA]\label{sutva}\textit{$Y_i = Y_i(A_i)$ for all study units. Further, each unit's potential outcomes are independent of the treatment status of any other unit: $Y_i(a) \indep A_j$ for all $i \neq j$}.
\end{assumption}
\vspace{-0.5em}
\begin{assumption}[Positivity]\label{positivity}
\textit{ $\PP (A=1 | \bX) \in (0,1)$.}
\end{assumption}
\vspace{-0.5em}
\begin{assumption}[Unconfoundedness]\label{unconfoundedness}
$(Y(1), Y(0)) \indep A | \bX$.
\end{assumption}
\vspace{-0.5em}
When Assumptions \ref{positivity}-\ref{unconfoundedness} additionally hold for all subjects with $S=1$, the treatment effect corresponding to the population of the validation data distribution, $\tau_\text{val} \eqdef \E(Y(1)-Y(0)|S=1)$, can be identified as
\begin{equation}
     \tau_\text{val} = \E_{\bX | S=1}[ \E (Y|\bX,S=1,A=1) - \E (Y|\bX,S=1,A=0)].
\end{equation}
The manner in which the validation data is obtained dictates whether $\tau_\val = \tau_\tate$. In particular, one common study design is to obtain a simple random sample from the main dataset to validate. Such study designs are consistent with an assumption that validation data is available completely at random:
\vspace{-0.5em}
\begin{assumption}[Validation completely at random]\label{srandom}
$(Y(1), Y(0), \bX, A, A^*) \indep S$.
\end{assumption} 
\vspace{-0.5em}
Under Assumption \ref{srandom}, we have $\tau_\text{TATE}=\tau_\text{val}$, implying that when Assumptions \ref{sutva}-\ref{unconfoundedness} also hold, one can consistently estimate $\tau_\text{TATE}$ using only observations from the validation data. While often employed in practice simple random validation samples are not always feasible. Instead, it is often the case that the availability of validation data is a function of observed, and potentially unobserved, baseline factors. To accommodate such settings, we  consider scenarios where the following assumptions hold in place of Assumption \ref{srandom}:
\stepcounter{assumption}
\begin{intassumption}[Covariate-dependent validation]\label{eff-mods}
 $(Y , A) \indep S | \bX$.
\end{intassumption}
\begin{intassumption}[Positivity of validation selection]\label{source-pos} $\PP(S=1|\bX) \in (0,1)$.   
\end{intassumption}
\ifthenelse{\equal{\showfig}{1}}{
\begin{figure}[h!]
    \centering
    \fontsize{16}{19}\selectfont
    \begin{subfigure}[b]{0.45\textwidth}
        \centering
        \resizebox{6cm}{4cm}{
            \begin{tikzpicture}[node distance=2cm]
                \node[circle,draw,minimum size=1.5cm] (Astar) at (6,-3) {$A^*$};
                \node[circle,draw,minimum size=1.5cm,] (X) at (0,0) {$X$};
                \node[circle,draw,minimum size=1.5cm,fill=lightgray] (A) at (3,0){$A$};
                \node[circle,draw,minimum size=1.5cm] (Y) at (6,0) {$Y$};
                \node[circle,draw,minimum size=1.5cm] (S) at (9,0) {$S$};
                \draw[->, thick, >=stealth] (A) -- (Astar);
                \draw[->, thick, >=stealth] (X) -- (A);
                \draw[->, thick, >=stealth] (A) -- (Y);
                \draw[->, dashed, thick, >=stealth] (Y) -- (Astar);
                \draw[->, thick, >=stealth] (X.north) to [out=50] (Y.north);
                \draw[->,dashed, thick, >=stealth] (X) -- (Astar);
                \draw [->, thick, >=stealth,opacity=0] (X.north) to [out=50] (S.north);
            \end{tikzpicture}
        }
        \caption{Selection into the validation data occurs completely at random.}
        \label{conf-dag-rand}
    \end{subfigure}
    \hfill
    \begin{subfigure}[b]{0.45\textwidth}
        \centering
        \resizebox{6cm}{4cm}{
            \begin{tikzpicture}[node distance=2cm]
                \node[circle,draw,minimum size=1.5cm] (Astar) at (6,-3) {$A^*$};
                \node[circle,draw,minimum size=1.5cm,] (X) at (0,0) {$X$};
                \node[circle,draw,minimum size=1.5cm,fill=lightgray] (A) at (3,0){$A$};
                \node[circle,draw,minimum size=1.5cm] (Y) at (6,0) {$Y$};
                \node[circle,draw,minimum size=1.5cm] (S) at (9,0) {$S$};
                \draw[->, thick, >=stealth] (A) -- (Astar);
                \draw[->, thick, >=stealth] (X) -- (A);
                \draw[->, thick, >=stealth] (A) -- (Y);
                \draw[->, thick, >=stealth] (X.north) to [out=50] (Y.north);
                \draw[->,dashed, thick, >=stealth] (X) -- (Astar);
                \draw[->,dashed, thick, >=stealth] (Y) -- (Astar);
                \draw [->, thick, >=stealth] (X.north) to [out=50] (S.north);
            \end{tikzpicture}
        }
        \caption{Selection into the validation data is random conditional on baseline covariates.}
        \label{conf-dag-genz}
    \end{subfigure}
     \begin{subfigure}[b]{0.45\textwidth}
        \centering
        \resizebox{6cm}{4cm}{
            \begin{tikzpicture}[node distance=2cm]
                \node[circle,draw,minimum size=1.5cm] (Astar) at (6,-3) {$A^*$};
                \node[circle,draw,minimum size=1.5cm,] (X) at (0,0) {$X$};
                \node[circle,draw,minimum size=1.5cm,fill=lightgray] (A) at (3,0){$A$};
                \node[circle,draw,minimum size=1.5cm] (Y) at (6,0) {$Y$};
                \node[circle,draw,minimum size=1.5cm] (S) at (9,0) {$S$};
                \draw[->, thick, >=stealth] (A) -- (Astar);
                \draw[->, thick, >=stealth] (X) -- (A);
                \draw[->, thick, >=stealth] (A) -- (Y);
                draw[->, thick, >=stealth] (Y) -- (S);
                \draw[->, thick, >=stealth] (X.north) to [out=50] (Y.north);
                \draw[->,dashed, thick, >=stealth] (X) -- (Astar);
                \draw[->,dashed, thick, >=stealth] (Y) -- (Astar);
                \draw [->, thick, >=stealth] (X.north) to [out=50] (S.north);
                \draw [->, thick, >=stealth] (Astar) -- (S);
                \draw [->, thick, >=stealth] (Y) -- (S);
            \end{tikzpicture}
        }
        \caption{Selection into the validation data is random conditional on all initially observed data.}
        \label{conf-dag-complex}
    \end{subfigure}
    \caption{Comparison of three scenarios involving classical, possibly differential measurement error. Dashed arrows indicate causal dependencies that can be present or absent. Our main methods consider scenarios consistent with the causal diagrams in panels \textbf{(a)} and \textbf{(b)}. 
    We consider scenarios consistent with panel \textbf{(c)} as an extension in Section \ref{sec:complex-val}.}
\end{figure} 
}

Under either set of independence assumptions, $\tau_\tate$ can be identified 
by the G-computation functional
\begin{equation}
\label{gen-functional}
\tau_\tate = \E_{\bX}[\E(Y|\bX,S=1,A=1) - \E(Y|\bX,S=1,A=0)].
\end{equation}  
Assumption \ref{eff-mods} guarantees that the conditional average treatment effect (CATE) function $\E(Y|\bX,A=a)=\E(Y|\bX,A=a,S=1)$, where the right-hand side is identifiable. One can then identify $\tau_\tate$ by marginalizing the CATE over the target population covariate distribution. Finally, to  leverage the error prone measurements $A^*$, we make assumptions analogous to the positivity and validation exchangeability conditions:
\stepcounter{assumption}
\begin{intassumption}[Conditional exchangeability of $A^*$ over $S$]\label{as:val-astar}
 $A^* \indep S | \bX$.
\end{intassumption}
\begin{intassumption}[Positivity of $A^*$]\label{as:astar-pos} $\PP(A^*=1|\bX) \in (0,1)$. 
\end{intassumption} 
The collective set of independence assumptions are consistent with the directed acyclic graphs presented in Figures \ref{conf-dag-rand} and \ref{conf-dag-genz}, and a proof of (\ref{gen-functional}) is provided in the Supplementary Materials. While Assumptions \ref{as:val-astar} and \ref{as:astar-pos} are not needed for \eqref{gen-functional} to hold, in the coming Section we demonstrate how these additional assumptions enable efficiency gains through our proposed method.
\vspace{-1.5em}
\section{Methods}
\label{methods}
\vspace{-1em}
\subsection{General Framework}
\label{cvmethod}
Originally developed by \cite{yang2019combining} to address situations with unmeasured confounding in observational studies, the approach later termed the \textit{control variates} method in \cite{guo2022multi} borrows variance reduction tools from the Monte Carlo sampling literature (\citealt{rubinstein1985efficiency}) and is applicable to scenarios where one has access to numerous sources of data, yet  the causal quantity of interest is only identifiable in a subset of those sources. We primarily focus on the setting where there are two sources, the validation data and the main data.
While previous applications have used the control variates approach to address partially missing confounders and selection bias, the method has not yet been adapted or evaluated in the context of error-prone exposure measurements, or measurement error in general. Further, existing applications of the control variates method  predominantly operate under the more restrictive independence Assumption \ref{srandom}, rather than allowing for the more general Assumption \ref{eff-mods} to hold. We extend the existing framework to account for both of these scenarios.
\\ \\
The control variates method is motivated by two key observations related to the estimation of $\tau_\tate$. First, notice that we can use the validated exposure measurements to construct an estimator $\hat \tau_\val$ of $\tau_\tate$ through the G-computation functional (\ref{gen-functional}), where the subscript indicates this estimator is primarily reliant on the validation data. For the moment we leave the form of $\hat \tau_\val$ unspecified. The resulting estimator, however, will likely be inefficient due to its heavy reliance on the typically small subset of observations with gold-standard exposure measurements. Second, consider an analogous G-computation functional that replaces $A$ with $A^*$,
\begin{equation}
\label{eq:control-var-genz}
     \E_{\bX} ( \E(Y | \bX, A^*=a,S=1)) = \E_{\bX} (\E(Y|\bX,A^*=a) ), \ a \in \{ 0,1\}.
\end{equation}
Notice that both sides of Equation (\ref{eq:control-var-genz}) are identified since $A^*$ is available for all subjects and equality holds by Assumptions \ref{sutva}-\ref{unconfoundedness} and \ref{eff-mods}-\ref{as:astar-pos}. While neither quantity represents a counterfactual parameter of interest, suppose we can construct consistent estimators for contrasts of the left- and right-hand sides of the above functionals, denoted $\hat \tau_\val^\ep$ and $\hat \tau_\main^\ep$ to emphasize that these are error-prone estimators of $\tau_\tate$, instead converging to some non-causal quantity $\tau^\ep$. Then notice  $\hat \tau_\val^\ep - \hat \tau_\main^\ep \cp 0$, where both estimators will in general be correlated with $\hat \tau_\val$. 
If we further suppose that these estimators satisfy

   \begin{equation}
        \sqrt{n} \begin{pmatrix}
        \hat \tau_\val - \tau_\text{TATE} \\ 
        \hat \tau^{\text{e.p.}}_\val - \hat \tau^{\text{e.p.}}_\main
    \end{pmatrix} \overset{D}{\rightarrow} N \left(\boldsymbol 0, \boldsymbol \Sigma \right), \ \ \ \boldsymbol \Sigma = \begin{pmatrix}
        v & \Gamma^\top \\ \Gamma & V
    \end{pmatrix},
    \label{asym-dist}
    \end{equation}
    
then we can consider a class of estimators for $\tau_\tate$ of the form $\hat \tau_\cv = \hat \tau_\val - b(\hat \tau_\val^\ep - \hat \tau_\main^\ep)$, where $b \in \mathbb{R}$. Setting $b =\Gamma^\top V^{-1}$ maximizes variance reduction and ensures that $\text{Var}(\hat \tau_\cv) \leq \text{Var}(\hat \tau_\val)$. Replacing $\Gamma$ and $V$ with their estimated values, for which we provide estimation details in Theorem \ref{thmgenz}, yields the proposed control variates estimator
\begin{equation}
\hat \tau_\cv = \hat \tau_\val - \hat \Gamma \hat V^{-1}(\hat \tau_\val^\ep - \hat \tau_\main^\ep).
\end{equation} \vspace{-0.25em}
The intuition behind the control variates method is to augment the initial consistent but inefficient estimator $\hat \tau_\val$ with an asymptotically mean zero variance reduction term $\hat \tau_\val^\ep - \hat \tau_\main^\ep$. This term is referred to as a control variate. Rather than directly model the measurement error mechanism, the control variates method leverages the correlation between these unbiased and error-prone estimators to yield an estimator with improved efficiency. While we focus on the scenario where the control variate is a scalar, in general the control variate can be multidimensional and based on estimators of other quantities so long as it is mean zero. 

In the remainder of this Section, we provide specific details on how to estimate each component comprising $\hat \tau_\cv$, with Figure \ref{cv-diagram} providing visual intuition. Throughout, we restrict our attention to implementations of the control variates method that make use of regular asymptotically linear (RAL) estimators, defined below.

\textbf{Definition 1}: An estimator $\hat \tau$ for a quantity $\tau$ is said to be \textit{asymptotically linear} if 
\[ \sqrt n (\hat \tau - \tau) = \frac{1}{\sqrt n} \sum_{i=1}^n \varphi(\bX_i) + o_\PP(1), \]
where $\varphi(\bX)$, which has zero mean and finite variance, is referred to as the \textit{influence function} for $\hat \tau$. An asymptotically linear estimator is RAL when it maintains the same asymptotic distribution under small perturbations to the data-generating probability distribution.  See \cite{van2000asymptotic} for a formal characterization of conditions ensuring regularity.
Two factors motivate this restriction: 1) many commonly-used causal effect estimators, such as augmented inverse probability weighting (AIPW) estimators, are RAL under modest conditions; and 2) the asymptotic distributions of RAL estimators tend to be more tractable than those of their non-RAL counterparts, making inference less cumbersome. 
\ifthenelse{\equal{\showfig}{1}}{
\begin{figure}[h!]
    \centering
    \includegraphics[width=\textwidth]{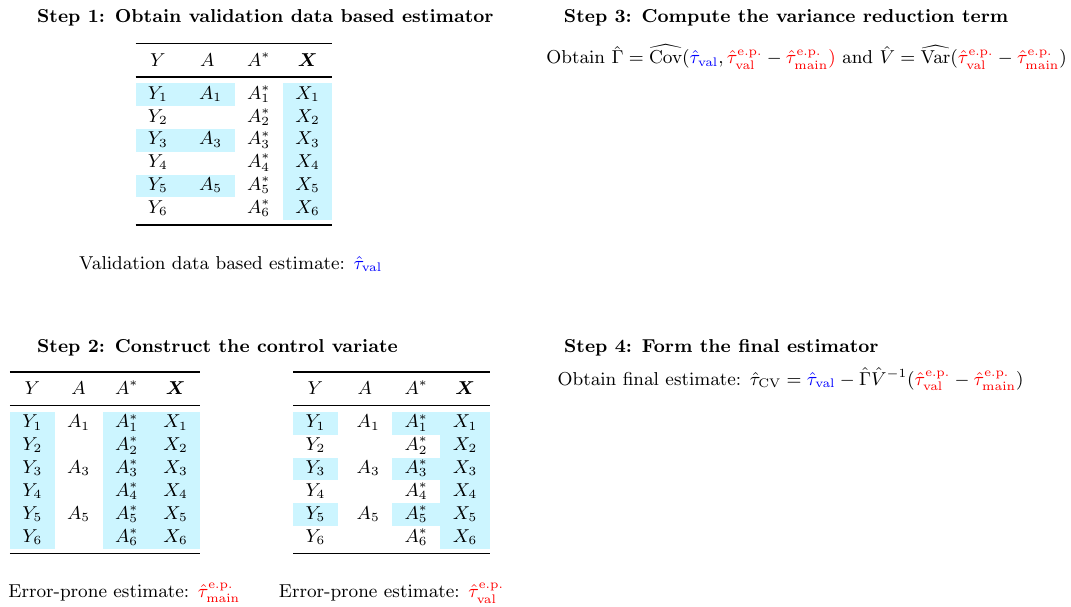}
    \caption{Illustration of the control variates method.}
    \label{cv-diagram}
\end{figure}
}

\subsection{Obtaining the Components of the Control Variates Estimator}
\label{obtaining-tauval}
To consistently estimate $\tau_\tate$ in this setting, we propose the use of doubly-robust estimators developed in the generalizability and transportability literature that leverage semiparametric efficiency theory. 
Specifically, we explore the efficient generalization estimators initially proposed by \cite{dahabreh2019generalizing}. 
These estimators leverage the observation that under Assumptions \ref{sutva}-\ref{unconfoundedness} and Assumptions \ref{eff-mods}-\ref{source-pos}, $\tau_\tate$ can be identified by the G-computation functional outlined in Equation (\ref{gen-functional}).
The aforementioned references have demonstrated the application of the efficient influence function derived from the functional displayed in Equation (\ref{gen-functional}), implying the doubly-robust estimator
\begin{equation}
\label{eq:val-dr}
 \mathsmaller{\hat \tau_\val = \frac{1}{n}\sum_{i=1}^n \left[ \left(\frac{A_i S_i}{\hat \kappa(\bX_i)\hat \pi(\bX_i)} - \frac{(1-A_i)S_i}{\hat \kappa(\bX_i)(1-\hat \pi(\bX_i))}
 \right) 
 \{Y_i-\hat \mu_{A_i}(\bX_i)\} 
 + \hat \mu_1(\bX_i) -  \hat \mu_0(\bX_i) \right]},    
\end{equation}
where $\mu_a(x) \eqdef \E(Y|\bX=x,A=a, S=1)$ and $\pi(x) \eqdef \PP (A=1|\bX=\bx, S=1)$ are estimated with the validation data, and $\kappa(x) \eqdef \PP(S=1|\bX=\bx)$ is estimated over the full sample. 
Notably, this approach does not require knowledge of the underlying measurement error model, and accommodates both Assumption \ref{srandom} and \ref{eff-mods}.

Similar to the construction of $\hat \tau_\val$, one can make an analogous adjustment to $\hat \tau_{\val}^{\text{e.p.}}$ to ensure the control variate is consistent for 0, by recalling equality of the two functionals displayed in Equation (\ref{eq:control-var-genz}). 
Recall that the key requirement for the control variate is not that this functional represents a causal effect, but rather that we can construct estimators in both the main and validation data that are both consistent for the same fixed quantity. Notice that the left-hand side of Equation (\ref{eq:control-var-genz}) can be consistently estimated using the same efficient generalization strategy described above with
\begin{equation}
\label{eq:ep-dr-val}
\hspace{-1em} \mathsmaller{\hat \tau_\val^\ep = \frac{1}{n} \sum_{i=1}^n  \hspace{-0.25em} \left[ \left(\frac{A_i^* S_i}{\hat \kappa(\bX_i)\hat \pi^\ep(\bX_i)} - \frac{(1-A_i^*)S_i}{\hat \kappa(\bX_i)(1-\hat \pi^\ep(\bX_i))}
\right)\hspace{-0.25em}
\{Y_i-\hat \mu^\ep_{A_i^*}(\bX_i)\}
+ \hat \mu^\ep_1(\bX_i) -  \hat \mu^\ep_0(\bX_i) \right]},
\end{equation}
where $\mu^\ep_a(x) \eqdef \hat \E(Y|\bX=x,A^*=a, S=1)$ and $\pi^\ep(\bx) \eqdef \hat \PP (A^*=1|\bX=\bx, S=1)$ are the error-prone CATE and propensity score models. The right-hand side of (\ref{eq:control-var-genz}) can be consistently estimated using the main data with the standard AIPW estimator
\begin{equation}
\label{eq:ep-dr}
\hspace{-1em} \mathsmaller{\hat \tau_\main^\ep = \frac{1}{n} \sum_{i=1}^n  \left[ \left(
\frac{A_i^*}{\hat g^\ep(\bX_i)} - \frac{1-A_i^*}{1-\hat g^\ep(\bX_i)}
\right)
\{
Y_i-\hat m^\ep_{A_i^*}(\bX_i)
\}
+ \hat m^\ep_1(\bX_i) -  \hat m^\ep_0(\bX_i) \right]},
\end{equation}
where $m^\ep_a(\bX) \eqdef \E[Y|A^*=a,\bX]$ and $g^\ep(\bX) \eqdef \PP(A^*=1|\bX)$ are error-prone analogues of the full-data CATE and propensity score functions $m_a(\bX)$ and $g(\bX)$ that do not condition on $S=1$. Since assumption \ref{as:val-astar} implies $\mu_a^\ep = m_a^\ep$, one can obtain an estimate of $\mu_a^\ep$ through regressing $Y$ on $A^*$ and $\bX$ in the validation data, or set $\hat \mu_a=\hat m_a$.
Importantly, we have not made any functional assumptions about the measurement error mechanism. 
\vspace{-1em}
\subsection{Theoretical Results}
\label{forming-cv}
\vspace{-0.25em}
Given a means for constructing each component of $\hat \tau_\cv$, we can turn our attention to inference. The following Theorem, whose proof is provided in the Supplementary Materials,  characterizes the asymptotic distribution of the control variates estimator (restricting consideration to RAL estimators) under Assumptions \ref{sutva}-\ref{unconfoundedness} and \ref{eff-mods}-\ref{source-pos}:
\vspace{-0.75em}
\begin{theorem}
\label{thmgenz}
Let $\hat \tau_\val$, $\hat \tau_\val^{\text{e.p.}}$ and $\hat \tau_\main^\ep$ be RAL estimators satisfying (\ref{asym-dist}) with corresponding influence functions $\varphi^*(A,\bX,Y), \phi^*(A^*,\bX,Y)$ and $\phi(A^*,\bX,Y)$,  respectively. For suitable estimators, under Assumptions \ref{sutva}-\ref{unconfoundedness} and Assumptions \ref{eff-mods}-\ref{source-pos} we have that 
$\sqrt{n} (\hat \tau_\cv -\tau_\tate)  \overset{D}{\rightarrow} N(0, v - \Gamma^2/V)$ where
\begin{flalign*}
& (1) \ v = \text{Var}(\varphi^*(A,\bX,Y)), \ \ \ \ \ \ \ \ \ \ (2) \  V = \text{Var}(\phi^*(A^*,\bX,Y) - \phi(A^*,\bX,Y)), \text{ and } & \\
& (3) \ \Gamma = \text{Cov}(\varphi^*(A,\bX,Y),\phi^*(A^*,\bX,Y) - \phi(A^*,\bX,Y)). &
\end{flalign*}
\end{theorem} 
\vspace{-1em}
There are multiple immediate consequences resulting from Theorem \ref{thmgenz}. First, notice that for a given set of RAL estimators used to construct $\hat \tau_\cv$,  $v, \Gamma$ and $V$ can be consistently estimated by their sample analogues, substituting in estimated values for the influence functions $\varphi^*(A,\bX,Y), \phi^*(A^*,\bX,Y)$ and $\phi(A^*,\bX,Y)$. We additionally provide a bootstrap procedure, similar to the one developed by \cite{guo2022multi}, in the Supplementary Materials. Second, the asymptotic normality of $\hat \tau_\cv$ enables straightforward inference and means for constructing confidence intervals, while also analytically quantifying the efficiency gain enjoyed from extracting information contained in $A^*$.
\vspace{-1em}
\subsubsection{Connections to Semiparametric Theory}

Recalling the observational unit $\bm O_i$, and letting $\bm O_i \sim \PP \in \mathcal{M}$ we briefly document efficiency and robustness properties of $\hat \tau_\cv$ when constructed through our recommended procedure in Section \ref{forming-cv}. Notably, while $\hat \tau_\val$ is an efficient RAL estimator of the generalization functional (\ref{gen-functional}) under Assumptions \ref{sutva}-\ref{unconfoundedness} and \ref{eff-mods}-\ref{source-pos}, $\hat \tau_\cv$ enables efficiency gains by additionally leveraging $A^*$ through Assumptions \ref{as:val-astar}-\ref{as:astar-pos}. In the Supplementary Materials, we show that our proposed $\hat \tau_\cv$ can be viewed as an RAL estimator in a semiparametric model which imposes  Assumptions \ref{as:val-astar}-\ref{as:astar-pos} on $\mathcal{M}$. The variance reduction term $\Gamma^2/V$ quantifies the efficiency gains this additional assumption allows through the control variates framework. 

A more subtle property of our proposed procedure is that when $\hat \tau_\val, \hat \tau_\val^\ep$ and $\hat \tau_\main^\ep$ are all obtained through  doubly-robust estimators, such as the ones outlined in Section \ref{obtaining-tauval}, then $\hat \tau_\cv$ will inherit the double robustness properties of its component estimators. 
Specifically, $\hat \tau_\cv$ will enjoy the consistency and, in our view more crucially, \textit{rate} robustness conditions of its underlying components.  This implies that if one wishes to make minimal assumptions about the underlying data-generating model and estimate all nuisance functions with  data-adaptive methods that themselves have slower rates of convergence, then the resulting control variates estimator can still achieve parametric $\sqrt n$ rates of convergence (\citealt{kennedy2024semiparametric}) under modest conditions attainable by many modern machine learning methods. 
We provide additional information on the precise robustness conditions of $\hat \tau_\cv$ in the Supplementary Materials.
\vspace{-1em}
\subsection{Complex Validation Data Sampling Schemes}
\label{sec:complex-val}
When researchers have control over which subjects to sample into the validation data, one may choose to adopt complex sampling rules that depend not only on the baseline covariates, but also the error-prone exposures $A^*$ and the observed outcomes $Y$. Biased sampling schemes have a long history in two-phase sampling designs (\citealt{breslow1999design,neyman1938contribution}). Relative to designs which validate subjects completely at random, biased sampling schemes can allow for efficiency gains by over-sampling observations that contribute higher degrees of information about the target estimand, particularly in scenarios where the exposure or outcome of interest is rare. Figure \ref{conf-dag-complex} displays a causal diagram consistent with such a biased sampling scheme.
\\ \\
The control variates method easily accommodates the above study design, provided the sampling mechanism into the validation data is known or can be estimated at a parametric rate.
Specifically, suppose selection into the validation data is determined according to a  sampling function $\kappa(\bX,A^*,Y) \eqdef \PP(S=1 | \bX, A^*, Y) \in (0,1)$, and Assumptions \ref{sutva}-\ref{unconfoundedness} hold. Notably, this sampling strategy invalidates Assumptions \ref{srandom}, \ref{eff-mods} and \ref{as:val-astar}. Since $Y$ is a direct cause of $S$ in this setting, we cannot rely on the conditional outcome mean invariance assumptions that drove our earlier proposed estimation procedure. In their place, we make the missing-at-random assumption $A \indep S | \bX, A^*, Y$ implied by this sampling strategy.
\\ \\
To form a control variates estimator, we propose the use of estimators based on weighted influence functions
\begin{align*}
\varphi^\text{IPSW}(\bm Z,A,S) &=
\frac{S}{ \kappa(\bm Z)}
\left\{
 m_1(\bX) - m_0(\bX) + \left(\frac{A}{ g(\bX)} - \frac{1-A}{1-g(\bX)} \right) m_{A}(\bX)
\right\},
\end{align*}
where $\bm Z = (\bX,A^*,Y)$. The above full-data nuisance functions $m_a$ and $g$,  defined in Section \ref{obtaining-tauval}, can be estimated through regression methods by adding weights $S/\kappa(\bm Z)$ to the underlying loss function (\citealt{rose2011targeted}). While the sampling probabilities $\kappa(\bm Z)$ will often be known in these settings, estimators which make use of a consistent estimator of $\kappa(\bm Z)$ will be more efficient \citep{tsiatis2006semiparametric}.
To construct a control variate, we consider functions of the form
\[
\phi^\text{IPSW,e.p.}(\bm Z,S) = \left(\frac{S}{\kappa(\bm Z)} -1 \right)\left\{
 m_1^\ep(\bX) - \hat m_0^\ep(\bX) + \left(\frac{A^*}{g^\ep(\bX)} - \frac{1-A^*}{1-g^\ep(\bX)} \right) m^\ep_{A^*}(\bX)
\right\}
\]
 noting  $\E[\phi^\text{IPSW,e.p.}(\bm Z,S)] =0$. Letting $\hat \tau^\text{IPSW}_\val = \frac{1}{n}\sum_{i=1}^n \hat \varphi^\text{IPSW}(\bm Z_i,A_i,S_i)$ and $\hat \phi^\text{IPSW,e.p.}_\main = \frac{1}{n}\sum_{i=1}^n \hat \phi^\text{IPSW,e.p.}(\bm Z_i,S_i)$,
the following Theorem summarizes how these two estimators can be used to construct a control variates estimator in this setting.
\begin{theorem}
\label{thm:complex-val}
Suppose  Assumptions \ref{sutva}-\ref{unconfoundedness} hold,  $A \indep S | \bm Z$, and $||\hat \kappa(\bm Z) - \kappa(\bm Z)|| = o_\PP(1/\sqrt n)$. Then, under additional regularity conditions outlined in the Supplementary Materials, we have  \vspace{-1.4em}
\begin{enumerate}
    \item $\hat \tau_\val^\text{IPSW}$ is asymptotically linear for $\tau_\tate$, with influence function $\varphi^\text{IPSW}(\bm Z,A,S) - \tau_\tate$, and 
    \item $\hat \phi_\main^\text{IPSW,e.p.}$ is asymptotically linear for $0$, with influence function $\phi^\text{IPSW}(\bm Z,S) - \tau^\ep$.
    \item 
    \begin{equation}
        \sqrt{n} \begin{pmatrix}
        \hat \tau_\val^\text{IPSW} - \tau_\text{TATE} \\ 
        \hat \phi^{\text{IPSW,e.p.}}_\main -0
    \end{pmatrix} \overset{D}{\rightarrow} N \left(\boldsymbol 0, \boldsymbol \Sigma \right), \ \ \ \boldsymbol \Sigma = \begin{pmatrix}
        v^\text{IPSW} & \Gamma^\text{IPSW} \\ \Gamma^\text{IPSW} & V^\text{IPSW}
    \end{pmatrix},
    \label{asym-dist-complex}
    \end{equation}
\end{enumerate}\vspace{-1.5em}
\begin{align*}
\Gamma^\text{IPSW}&=\text{Cov}\left( \phi^\text{IPSW,e.p.}(\bm Z, S), \varphi^\text{IPSW}(\bm Z, A,S)\right); \\
V^\text{IPSW} &= \text{Var}\left(  \phi^\text{IPSW,e.p.}(\bm Z, S) \right).
\end{align*} 
\end{theorem}
The proof of Theorem \ref{thm:complex-val} is provided in the Supplementary Materials. A key distinction with the estimator proposed in Section \ref{cvmethod} is that control variate is ensured to be mean zero through the multiplicative factor $(S/\kappa(\bm Z) -1)$, rather than through a difference in error-prone estimators. This distinction is necessary, as the originally proposed estimator $\hat \tau_\val^\ep$ is no longer consistent for $\tau^\ep$ in this broader sampling scheme. An analogous construction was explored in \cite{yang2019combining} for partially missing covariates.
\\ \\
Our result extends findings in \cite{yang2019combining} by relaxing the requirement that the true CATE and propensity score functions lay in parametric modeling classes. While we require faster parametric $\sqrt n$ consistent estimation of $\kappa(\bm Z)$, such rates are attainable in settings where the true probabilities are controlled by design, but estimated for efficiency gains. An immediate corollary of Theorem \ref{thm:complex-val} is that control variates estimators of the form $ \hat \tau_\cv = \hat \tau_\val^\text{IPSW} - (\hat \Gamma^\text{IPSW}/\hat V^\text{IPSW}) \hat \phi_\main^\text{IPSW,e.p.},$ where $\hat \tau_\main^\ep$ is obtained as in (\ref{eq:ep-dr}), will be consistent for $\tau_\tate$. Further, the asymptotic linearity of all three component estimators implies one can estimate $\Gamma^\text{IPSW}$ and $V^\text{IPSW}$ through the influence functions of each estimator.
\vspace{-1em}
\section{Simulation Study}
\label{simmy}
\vspace{-1em}
\subsection{Setting}
\label{sec:simmy-setting}
To investigate the performance of the control variates method in finite-sample settings, we conducted an extensive simulation exercise. We implemented the control variates method with the doubly-robust component estimators for $\hat \tau_\val, \hat \tau_\val^\ep$ and $\hat \tau_\main^\ep$ outlined in Section \ref{obtaining-tauval}.  We compared the control variates estimator to an oracle (best case) AIPW estimator where one has access to the true $A$ for every observation, a na\"ive AIPW estimator that ignores measurement error and uses $A^*$ in place of $A$, the generalization estimator outlined in (\ref{eq:val-dr}), a validation data only AIPW estimator and multiple imputation for measurement error (denoted MIME). MIME is a  standard approach taken for addressing measurement error with validation data (\citealt{webb2017imputation,josey2023estimating}) and has been compared to control variates estimators in missing covariate settings (\citealt{yang2019combining}). We emphasize that the generalization estimator (\ref{eq:val-dr}) $\hat \tau_\val$ is the initial consistent estimator used in forming the control variates estimator, whose variance is reduced by the inclusion of the control variate term. For each estimator, we estimated all underlying nuisance models with a Super Learner (\citealt{van2007super}).
Full details are provided in the Supplementary Materials. 
\\ \\
We consider two general scenarios: 1) the validation samples are selected completely at random, and 2) the validation samples are selected conditionally as a function, which we treat as unknown, of the measured covariates $\bX$. Within each scenario, we examined how the competing estimators perform under varying levels of measurement error severity and by altering the relative sizes of the validation data, $\rho \eqdef \PP(S=1)$.
We generated $5{,}000$ datasets of $n=5{,}000$ observations through the following process:
\begin{align*}
\bX_i &\sim N(\bm 1, \bm \Sigma_{\bX})
& (\text{Covariates}); \\
A_i | \bX_i &\sim \text{Bernoulli}(\pi(\bX_i)), \enskip \pi(\bX_i) = \text{expit}(\alpha_0 + \bX_i^\top \bm \alpha) & (\text{Exposure}); \\
A_i^* | A_i &\sim \text{Bernoulli}(p_i),  \ p_i = A_i \delta + (1-A_i)\zeta & (\text{Measurement}); \\
S_i | \bX_i &\sim \text{Bernoulli}(\kappa(\bX_i)), \enskip \kappa(\bX_i) = \frac{\rho \cdot \text{expit}(\eta_0 + \bm X_i^\top \bm \eta)}{\frac{1}{n}\sum_{k=1}^{n} \text{expit}(\eta_0 + \bm X_k^\top \bm \eta)} & (\text{Val. data selection}); \\
Y_i | A_i, \bX_i &\sim N(\mu(\bX_i), \varepsilon), \enskip \mu(\bX_i) = \beta_0 + \tau A_i + \bX_i^\top \bm \beta + A_i\bX_i^\top \bm \gamma & (\text{Outcome}). 
\end{align*}
We performed multiple imputation using the \texttt{mice} package in R (\citealt{van2011mice}), generating $10$ imputed datasets through predictive mean matching (\citealt{little1988missing}) and implementing Rubin's combining rules to obtain our final treatment effect and variance estimates. 
Predictive mean matching is the default imputation model option in \texttt{mice}, and has been demonstrated to flexibly account for complex missing data patterns (\citealt{kleinke2017multiple}). We implement the control variates method with the \texttt{controlVariatesME}
R package, which we developed to facilitate use of our proposed methods. In implementing the control variates method, we use the asymptotic expressions from Theorem \ref{thmgenz} to estimate $\Gamma$ and $V$. Full details on the simulation design and replication code are included in the Supplementary Material. Additional simulation exercises altering the overall sample size, specification of nuisance learners, and the sampling setting considered in Section 3.4 can be found in the Supplementary Materials, where our qualitative findings are similar to those discussed in the coming section.

\vspace{-1.3em}
\subsection{Results}
Figures \ref{bias-plot} and \ref{rmse-plot} display results from the simulation  in settings where the validation data is obtained completely at random and conditionally at random given $\bX$, respectively.
We report the percent bias, 95\% confidence interval coverage rate, and root mean square error (RMSE) of each estimator under both sampling scenarios, against varying sensitivity levels $\delta$---a measure of measurement error severity ---and relative sizes of the validation data. 
\ifthenelse{\equal{\showfig}{1}}{
\begin{figure}[h!]
    \centering
    \includegraphics[scale=.7]{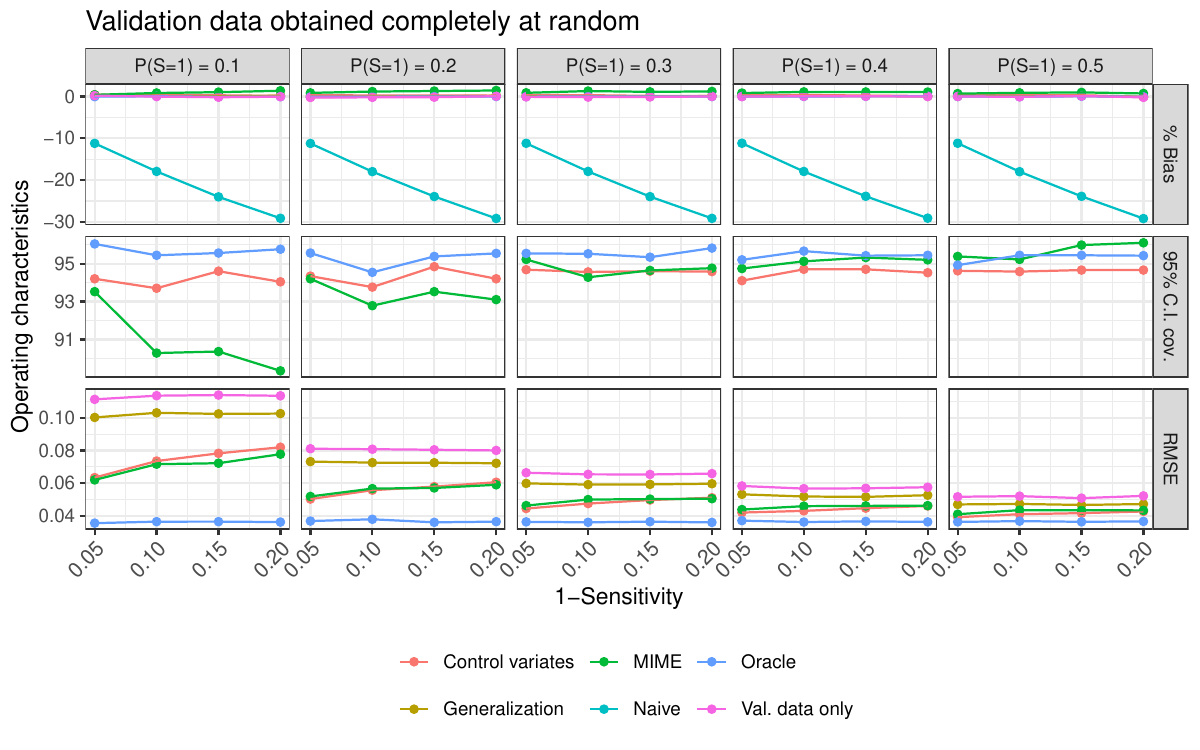}
    \caption{Percent bias, 95\% CI coverage and RMSE of each method by sensitivity, relative size of validation data to main data. Validation data is obtained completely at random. RMSE is only displayed for estimators that are not severely biased, to avoid distorting the scale. Simulation results are averaged  over $5000$ iterations, fixing $n_\main = 5000$.}
    \label{bias-plot}
\end{figure}
\begin{figure}[h!]
    \centering
    \includegraphics[scale=.7]{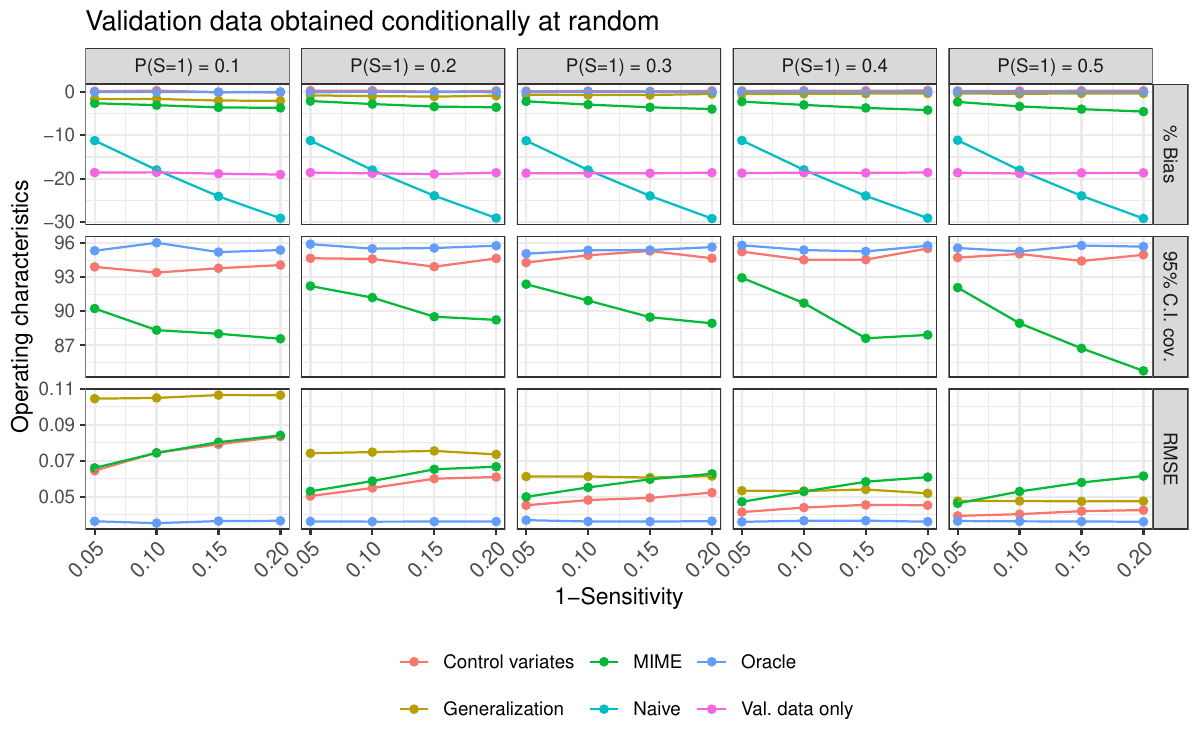}
    \caption{Percent bias, 95\% CI coverage and RMSE of each method by sensitivity and relative size of validation data to main data. Validation data is obtained conditionally on $\bX$, inducing covariate shift between the main and validation data. RMSE is only displayed for estimators that are not severely biased, to avoid distorting the scale. Simulation results are  averaged over $5000$ iterations, fixing $n_\main = 5000$.}
    \label{rmse-plot}
\end{figure}} 

Beginning with Figure \ref{bias-plot}, there are three key takeaways. First, we see that in all settings, the na\"ive estimator is considerably biased, highlighting the general need to account for exposure measurement error. Second, we see that multiple imputation and the control variates method both offer substantial RMSE reduction relative to the validation data only estimator across increasingly severe measurement error, regardless of the validation data size, but particularly as smaller portions of the main sample are included in the validation sample. Finally, due to a minor mis-specification in the predictive mean matching imputation model, multiple imputation is slightly biased, leading to  undercoverage of the nominal 95\% confidence level.
\\[1.1em]
In Figure \ref{rmse-plot}, where the validation data is obtained at random but inclusion occurs conditionally on the covariates, there are again three key takeaways. First, na\"ively using a validation data only estimator, which ignores covariate shift between the main and validation data, leads to substantial bias. Second,  multiple imputation exhibits minor degrees of bias that hampers its coverage rates. Intuitively, the introduction of covariate shift further complicates the underlying true---but unknown---imputation model, with the resulting misspecification of a predictive mean matching approach propagating into the final estimate. Finally, we see that the control variates estimator is again unbiased but in this scenario it outperforms MIME in terms of RMSE due to the misspecification bias of the latter estimator under this validation sampling scenario.

\section{Data Example}
\label{application}

We applied the control variates method to EHR data from the VCCC introduced in Section \ref{sec:intro}. The VCCC data has been featured in several works focusing on measurement error-correction methods, including \cite{oh2021raking,giganti2020accounting} and \cite{amorim2021two}. 
Data were collected on numerous baseline characteristics at each patient's initial visit, as well as clinical data at all follow-up visits. 
Continuous characteristics, denoted $\bm C_i$, included age at first visit, while discrete characteristics $\bX_i$ included risk factors such as injection drug use, and demographic characteristics including sex, race and ethnicity.
See the Supplementary Materials for additional information on all relevant variables.

The VCCC validated error-prone records for all 4{,}217 patients, resulting in an initial unvalidated dataset and a corresponding fully-validated dataset while revealing considerable error in numerous EHR-derived variables. Among the severely error-prone variables was 
the  occurrence of an AIDS-defining event (ADE) prior to first visit, an indicator of severely delayed initiation of treatment. Access to the  complete set of both validated and error-prone ADE measurements provides an ideal scenario for examining the real-world performance of the control variates method relative to na\"ive and validation-data only estimators. 

To evaluate the control variates method, we emulate a scenario in which a researcher seeks to estimate the average causal effect of having suffered an ADE at baseline ($A$), which possesses initial error-prone measurements $A^*$, on the 5-year post-baseline risk of death ($Y$) among patients with no history of antiretroviral therapy (ART) use prior to initiating care with the VCCC. 
Using the validated ART data to exclude patients who initiated ART prior to enrollment at the VCCC---a common exclusion criterion in HIV studies---1{,}907 patients remained. The misclassification rate of ADE among these 1{,}907 patients was 0.096, where much of this misclassification was due to false positives. Notably, the error-prone ADE indicator exhibited a false positive rate of 0.420, with a more modest false negative rate of 0.031. Given that pre-baseline ADEs were relatively infrequent in the validated data, with a prevalence of 0.123, this suggests that na\"ively using the error-prone ADE indicators can result in substantial bias.
\\ \\
With access to validated ADE indicators for every patient, we can examine the performance of the control variates estimator under different relative sizes of the validation data by artificially ignoring varying proportions of validation data, pretending we do not have access to the remaining validated ADE indicators. To do this, we considered the same set of relative sizes as in the simulation exercise. At each validation size, we created 1{,}000 hypothetical internal validation datasets via simple random sampling without replacement from the original, full validation dataset. To ensure we know the true underlying causal parameter, we simulated a synthetic 5-year survival outcome by (1) fitting a logistic regression model
\begin{equation}
  \E[Y_i|A_i, \bX_i, \bm C_i] = \text{expit}(\alpha A_i + A_i \bX_i^\top \bm \gamma + \bm C_i^\top \bm \beta) 
  \label{synth-outcome-mod}
\end{equation}
\noindent using the validated ADE measurements and (2) at each simulation iteration, used the fitted model (\ref{synth-outcome-mod}) to a generate a new realization of the synthetic outcome so that each of the 1{,}000 simulated internal validation datasets has an accompanying synthetic survival outcome.
We implemented the control variates method using each of the hypothetical validation datasets, with $\hat \tau_\val$, $\hat \tau_\val^\ep$ and $\hat \tau_\main^\ep$ estimated via the methods outlined in Section \ref{cvmethod}. We compare the control variates estimator to the generalization estimator $\hat \tau_\val$, as well as the same na\"ive and oracle AIPW estimators outlined in Section \ref{simmy}. All nuisance models were estimated with a Super Learner 
that adjusts for the baseline factors $\bX_i$ and $\bm C_i$. See the Supplementary Materials  for further details on our implementation and Super Learner libraries.
\ifthenelse{\equal{\showfig}{1}}{
\begin{figure}[h!]
    \centering
    \includegraphics[scale=0.7]{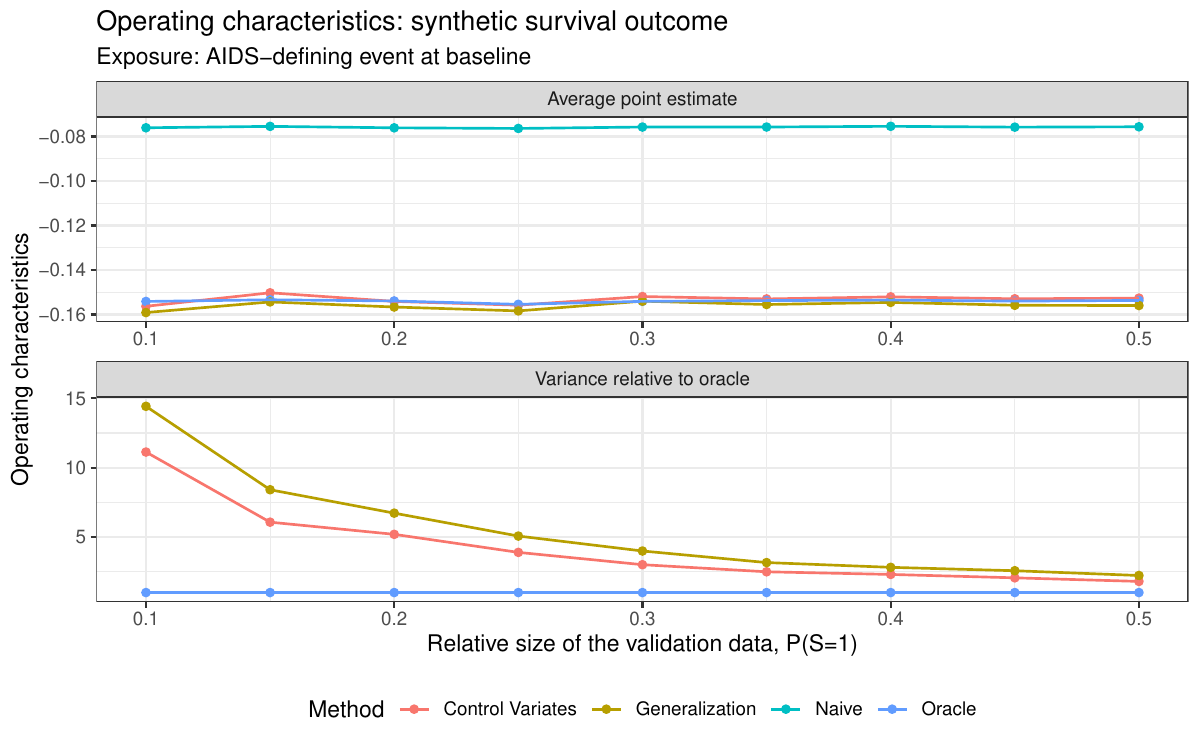}
    \caption{Results from the VCCC data application. }
    \label{vccc-results}
\end{figure}
}

The results of our analysis are displayed in Figure \ref{vccc-results}. We present the average point estimate of each method and their relative efficiency compared to the oracle estimator at each value of $\rho$ considered.
Similar to the findings from the simulation study, we see that ignoring measurement error generates substantial bias---on the order of 50\% in this example---while the generalization and control variates estimators are unbiased by design. We also observe that the control variates estimator provides considerable variance reduction relative to the validation data based estimator, improving the efficiency of the validation data based estimator by roughly 20\% across all validation sizes considered. This reduction in variance is notable since exposure measurement validation will tend to be costly in practice. Our results suggests that by applying our proposed methods, one can attain efficiency levels that would otherwise require additional expensive data validation.
\section{Discussion}
\label{disc}
Research at the intersection of measurement error and causal inference is a relatively new endeavor. Nevertheless, there is a growing need for flexible, modern methods for the estimation of causal effects that can accommodate various study designs while addressing measurement error. In this paper, we make contributions to this area of research by introducing the control variates estimator as a means for addressing exposure measurement error, under two-phase sampling data collection processes. 
Theory and our simulation studies show that relative to multiple imputation, our proposed method is more robust to model misspecification, as multiple imputation requires consistent estimation of the exposure imputation model while the control variates method does not require one to specify an imputation model.  We also demonstrate the flexibility of our proposed approach, showing its ability to address  complex validation data sampling schemes. Using recent developments from the generalizability and transportability literature, we extended earlier work by \cite{yang2019combining}, deriving the asymptotic distribution of a generic control variates estimator that makes use of efficient generalization and transportation estimation procedures. Theory, our simulation study and data application all demonstrate that the control variates method can substantially reduce the number of validation samples required to achieve a desired level of efficiency.
\\ \\
There are many avenues for future study mentioned intermittently throughout this article. While the simulation study provides preliminary insights into the performance of the control variates estimator relative to common approaches to measurement error correction, it would be valuable to consider more complicated data-generating processes and, in particular, a more exhaustive set of measurement error mechanisms. Further, while the focus of this paper is on measurement error in the exposure of interest, the control variates method naturally extends to scenarios where there is measurement error in the outcome, or scenarios where the outcome is partially missing with full information on surrogate outcomes. Considerations of this extension and comparison with methods like multiple imputation and the method described in \cite{kallus2025role}, would be valuable for scientists trying to discern the best method for their analytical challenges.

\subsubsection*{Acknowledgements}
This work was funded by the National Institute of Health (NIH) grants T32AI007358, K01ES032458, R37AI131771 and P30AI110527.

\subsubsection*{Data Availability Statement}

The data utilized in this study were obtained from Vanderbilt University Medical Center under a data use agreement (DUA) and are not publicly available. Access to the data is subject to approval by Vanderbilt University Medical Center and may be requested  through direct inquiry to the institution.


\fancyhf{}  
\fancyfoot[C]{\thepage}  

\bibliographystyle{biom}
\bibliography{sources}


\clearpage

\ifthenelse{\equal{\showfig}{0}}{
\begin{figure}[h!]
    \centering
    \fontsize{16}{19}\selectfont
    \begin{subfigure}[b]{0.45\textwidth}
        \centering
        \resizebox{6cm}{4cm}{
            \begin{tikzpicture}[node distance=2cm]
                \node[circle,draw,minimum size=1.5cm] (Astar) at (6,-3) {$A^*$};
                \node[circle,draw,minimum size=1.5cm,] (X) at (0,0) {$X$};
                \node[circle,draw,minimum size=1.5cm,fill=lightgray] (A) at (3,0){$A$};
                \node[circle,draw,minimum size=1.5cm] (Y) at (6,0) {$Y$};
                \node[circle,draw,minimum size=1.5cm] (S) at (9,0) {$S$};
                \draw[->, thick, >=stealth] (A) -- (Astar);
                \draw[->, thick, >=stealth] (X) -- (A);
                \draw[->, thick, >=stealth] (A) -- (Y);
                \draw[->, dashed, thick, >=stealth] (Y) -- (Astar);
                \draw[->, thick, >=stealth] (X.north) to [out=50] (Y.north);
                \draw[->,dashed, thick, >=stealth] (X) -- (Astar);
                \draw [->, thick, >=stealth,opacity=0] (X.north) to [out=50] (S.north);
            \end{tikzpicture}
        }
        \caption{Selection into the validation data occurs completely at random.}
        \label{conf-dag-rand}
    \end{subfigure}
    \hfill
    \begin{subfigure}[b]{0.45\textwidth}
        \centering
        \resizebox{6cm}{4cm}{
            \begin{tikzpicture}[node distance=2cm]
                \node[circle,draw,minimum size=1.5cm] (Astar) at (6,-3) {$A^*$};
                \node[circle,draw,minimum size=1.5cm,] (X) at (0,0) {$X$};
                \node[circle,draw,minimum size=1.5cm,fill=lightgray] (A) at (3,0){$A$};
                \node[circle,draw,minimum size=1.5cm] (Y) at (6,0) {$Y$};
                \node[circle,draw,minimum size=1.5cm] (S) at (9,0) {$S$};
                \draw[->, thick, >=stealth] (A) -- (Astar);
                \draw[->, thick, >=stealth] (X) -- (A);
                \draw[->, thick, >=stealth] (A) -- (Y);
                \draw[->, thick, >=stealth] (X.north) to [out=50] (Y.north);
                \draw[->,dashed, thick, >=stealth] (X) -- (Astar);
                \draw[->,dashed, thick, >=stealth] (Y) -- (Astar);
                \draw [->, thick, >=stealth] (X.north) to [out=50] (S.north);
            \end{tikzpicture}
        }
        \caption{Selection into the validation data is random conditional on baseline covariates.}
        \label{conf-dag-genz}
    \end{subfigure}
     \begin{subfigure}[b]{0.45\textwidth}
        \centering
        \resizebox{6cm}{4cm}{
            \begin{tikzpicture}[node distance=2cm]
                \node[circle,draw,minimum size=1.5cm] (Astar) at (6,-3) {$A^*$};
                \node[circle,draw,minimum size=1.5cm,] (X) at (0,0) {$X$};
                \node[circle,draw,minimum size=1.5cm,fill=lightgray] (A) at (3,0){$A$};
                \node[circle,draw,minimum size=1.5cm] (Y) at (6,0) {$Y$};
                \node[circle,draw,minimum size=1.5cm] (S) at (9,0) {$S$};
                \draw[->, thick, >=stealth] (A) -- (Astar);
                \draw[->, thick, >=stealth] (X) -- (A);
                \draw[->, thick, >=stealth] (A) -- (Y);
                draw[->, thick, >=stealth] (Y) -- (S);
                \draw[->, thick, >=stealth] (X.north) to [out=50] (Y.north);
                \draw[->,dashed, thick, >=stealth] (X) -- (Astar);
                \draw[->,dashed, thick, >=stealth] (Y) -- (Astar);
                \draw [->, thick, >=stealth] (X.north) to [out=50] (S.north);
                \draw [->, thick, >=stealth] (Astar) -- (S);
                \draw [->, thick, >=stealth] (Y) -- (S);
            \end{tikzpicture}
        }
        \caption{Selection into the validation data is random conditional on all initially observed data.}
        \label{conf-dag-complex}
    \end{subfigure}
    \caption{Comparison of three scenarios involving classical, possibly differential measurement error. Dashed arrows indicate causal dependencies that can be present or absent. Our main methods consider scenarios consistent with the causal diagrams in panels \textbf{(a)} and \textbf{(b)}. 
    We consider scenarios consistent with panel \textbf{(c)} as an extension in Section \ref{sec:complex-val}.}
\end{figure}

\begin{figure}[h!]
    \centering
    \includegraphics[width=\textwidth]{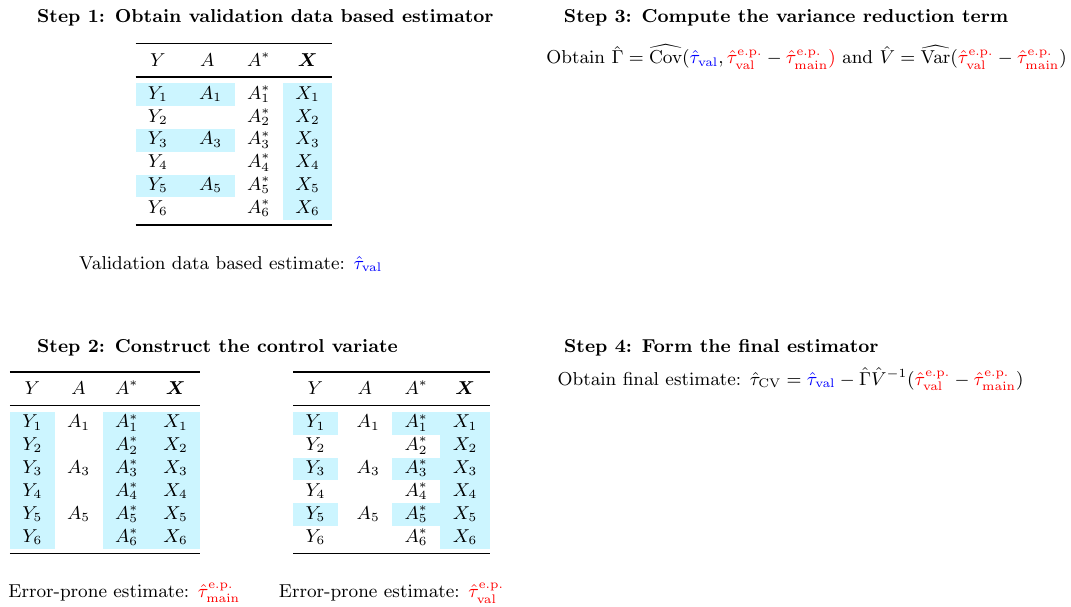}
    \caption{Illustration of the control variates method.}
    \label{cv-diagram}
\end{figure}

\begin{figure}[h!]
    \centering
    \includegraphics[scale=.7]{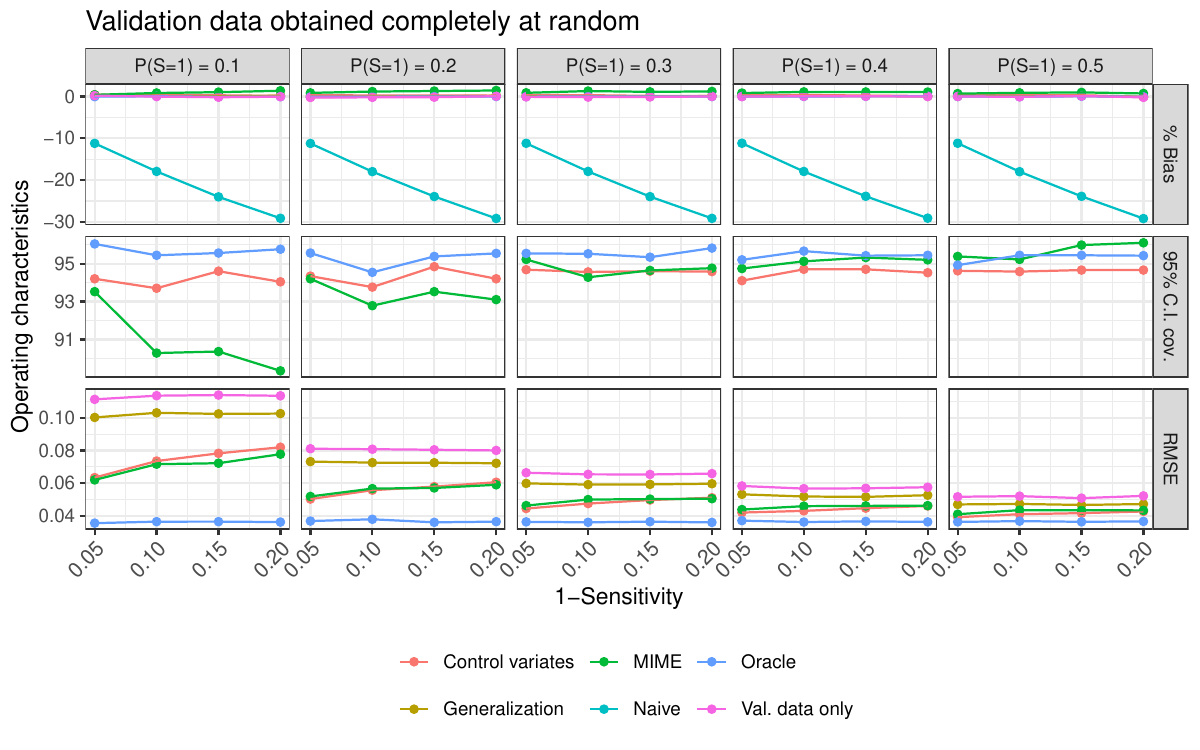}
    \caption{Percent bias, 95\% CI coverage and RMSE of each method by sensitivity, relative size of validation data to main data. Validation data is obtained completely at random. RMSE is only displayed for estimators that are not severely biased, to avoid distorting the scale. Simulation results are averaged  over $5000$ iterations, fixing $n_\main = 5000$.}
    \label{bias-plot}
\end{figure}
\begin{figure}[h!]
    \centering
    \includegraphics[scale=.7]{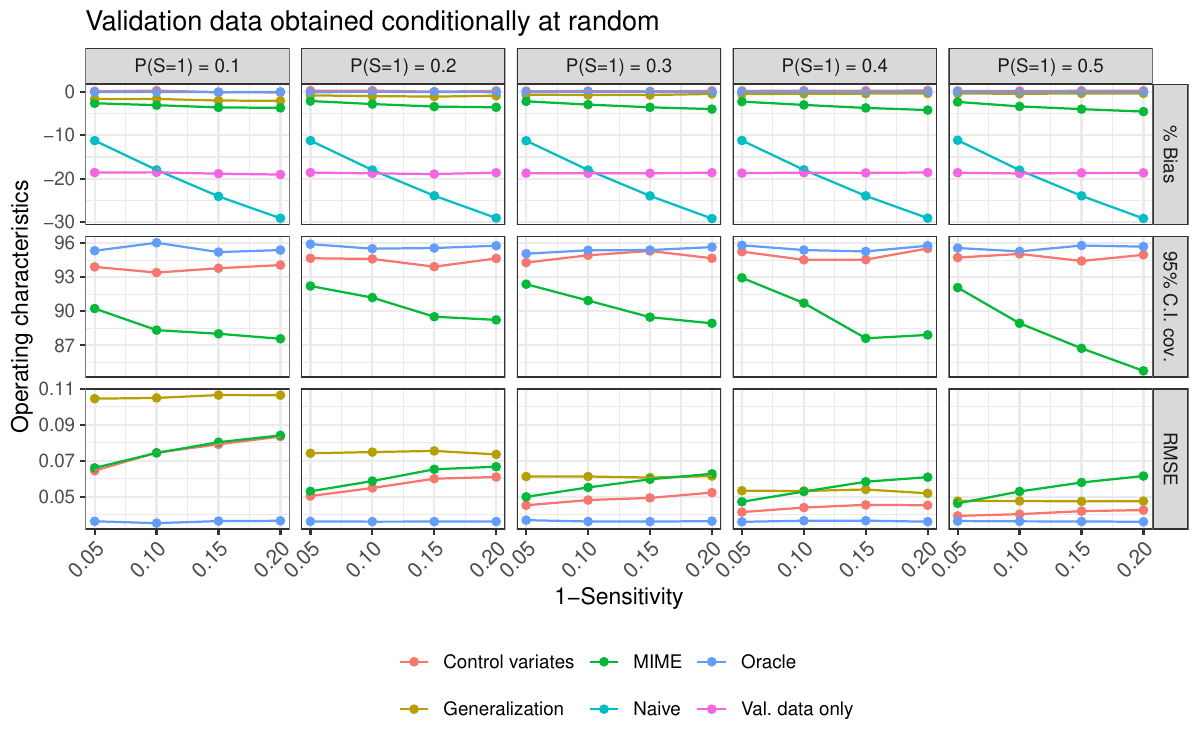}
    \caption{Percent bias, 95\% CI coverage and RMSE of each method by sensitivity and relative size of validation data to main data. Validation data is obtained conditionally on $\bX$, inducing covariate shift between the main and validation data. RMSE is only displayed for estimators that are not severely biased, to avoid distorting the scale. Simulation results are  averaged over $5000$ iterations, fixing $n_\main = 5000$.}
    \label{rmse-plot}
\end{figure}

\begin{figure}[h!]
    \centering
    \includegraphics[scale=0.7]{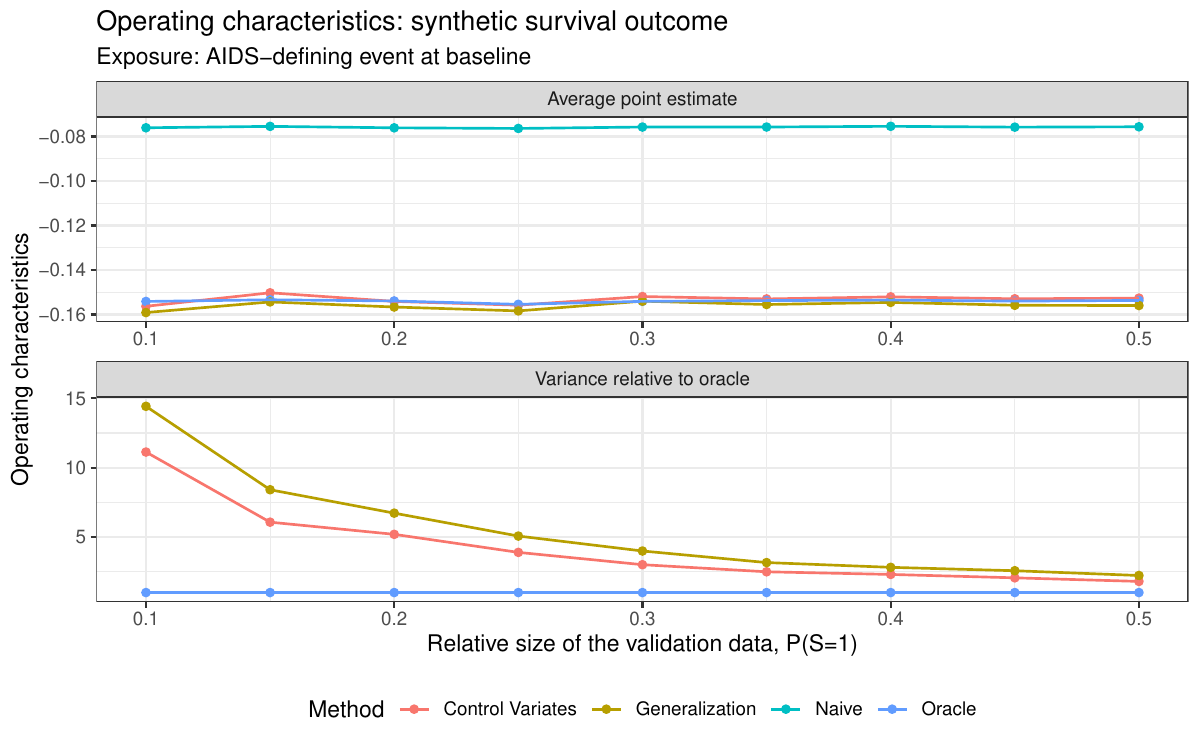}
    \caption{Results from the VCCC data application. }
    \label{vccc-results}
\end{figure}
}

\clearpage


\appendix

\setcounter{equation}{0}
\setcounter{table}{0}
\renewcommand{\theequation}{A\arabic{equation}}

\renewcommand{\thetable}{A\arabic{table}}


\renewcommand{\thefigure}{A\arabic{figure}}
\captionsetup[figure]{labelformat=webfig, labelsep=colon}
\captionsetup[table]{labelformat=webtab, labelsep=colon}
\setcounter{figure}{0}
\setcounter{table}{0}

\setstretch{1.5}
\begin{center}
\large{Supplementary Material for ``Flexible and Efficient Estimation of Causal Effects with Error-Prone Exposures: A Control Variates Approach for Measurement Error"
\\ by K. Barnatchez,  R. Nethery, B. Shepherd, G. Parmigiani, and  K. Josey}
\end{center}

\section*{Web Appendix A: Proofs}
\label{proofs}

\subsection*{Proof of Theorem \ref{thmgenz}}
Since $\hat \tau_\val, \hat \tau_\val^\ep$ and $\hat \tau_\main^\ep$ are all RAL, we have
\begin{align*}
    \sqrt{n} (\hat \tau_\val - \tau_\tate) &= \frac{1}{\sqrt{n}} \sum_{i=1}^{n} \varphi^*(A_i,\bX_i,Y_i,S_i) + o_\PP(1); \\
    \sqrt{n} (\hat \tau^\ep_\main - \tau_\text{e.p.}) &=  \frac{1}{\sqrt{n}}\sum_{i=1}^{n} \phi(A_i^*, \bX_i, Y_i) + o_\PP(1); \\
    \sqrt{n} (\hat \tau^{\text{e.p.}}_\val - \tau_\text{e.p.}) &=  \frac{1}{\sqrt{n}}\sum_{i=1}^{n} \phi^*(A_i^*, \bX_i, Y_i,S_i) + o_\PP(1).
\end{align*}
This implies
\begin{align*}
   \sqrt{n} (\hat \tau^{\text{e.p.}}_\val - \hat \tau^\ep_\main) &=  \frac{1}{\sqrt{n}}\sum_{i=1}^{n} \left[\phi^*(A_i^*, \bX_i, Y_i) -  \phi(A_i^*, \bX_i, Y_i)\right].
\end{align*}
Since all observations are i.i.d. we have that   
\begin{align*}
    \Var \left( \frac{1}{\sqrt{n}}\sum_{i=1}^{n} \phi^*(A_i^*, \bX_i, Y_i) \right) &= \Var (\phi^*(A^*, \bX, Y)) = v, \text{ and} \\
    \Var \left( \frac{1}{\sqrt{n}}\sum_{i=1}^{n} \left[\phi^*(A_i^*, \bX_i, Y_i) -  \phi(A_i^*, \bX_i, Y_i)\right]\right) &=  \Var (\phi^*(A^*, \bX, Y) - \phi(A^*, \bX, Y)) = V,
\end{align*}
Through similar reasoning,
\begin{align*}
 \Cov\left(    \sqrt{n} (\hat \tau_\val - \tau_\tate), \sqrt{n} (\hat \tau_\val^\ep - \hat \tau^\ep_\main)  \right) &= \Cov (\varphi^*(A_i,\bX_i,Y_i,S_i), \phi^*(A^*, \bX, Y) - \phi(A^*, \bX, Y)) \\
 &= \Gamma
\end{align*}
Let $\Gamma$ and $v$ be estimated by their sample analogues. Recalling $\hat \tau_\cv = \hat \tau_\val - \hat \Gamma / \hat V (\hat \tau_\val^\ep - \hat \tau_\main^\ep)$, and that all 3 component estimators are RAL, asymptotic normality directly follows by Slutsky's Theorem. To establish the asymptotic variance of $\hat \tau_\cv$, notice that asymptotically
\begin{align*}
   \text{Var}(\hat \tau_\cv) &= \text{Var}(\hat \tau_\val) + (\Gamma /V)^2\text{Var}(\hat \tau_\val^\ep - \hat \tau_\main^\ep) - 2\Gamma/V\text{Cov}(\hat \tau_\val, (\hat \tau_\val^\ep - \hat \tau_\main^\ep)) \\
   &=
 \text{Var}(\hat \tau_\val) + \frac{1}{n}\Gamma^2/V - \frac{1}{n}2\Gamma^2/V \\
 &=
 \frac{1}{n}\left(v - \Gamma^2/V\right).
\end{align*}

\subsection*{Connection of $\hat \tau_\cv$ to Semiparametric Theory}

Recall that $O_i = (Y_i,A_i,A_i^*,\bX_i,S_i) \sim \PP \in \mathcal{M}$.
While $\hat \tau_\val$, presented in (1), is an efficient nonparametric RAL estimator of the generalization functional within a model in which only Assumptions \ref{sutva}-\ref{unconfoundedness} and Assumptions \ref{eff-mods}-\ref{source-pos} hold, the results of Theorem \ref{thmgenz} imply $\hat \tau_\cv$ can  be viewed as an RAL estimator of the generalization functional in a model which places the restriction $Y \indep R | \bX, A^*$ on $\mathcal{M}$, since the additional Assumption \ref{as:val-astar} implies this latter independence.
\\ \\
In such a restricted model, the proposed $\hat \tau_\cv$ is just one of many possible RAL estimators. To derive the RAL estimator with lowest variance in this restricted model, one would need to characterize the closed linear span of all possible parametric submodels that satisfy this restriction,
and then subtract off the projection of any RAL estimator, such as $\hat \tau_\val$ or $\hat \tau_\cv$ onto the orthogonal complement of this subspace (\cite{van2000asymptotic}). Such a derivation is beyond the scope of this work. Further, a major strength of the control variates method is its ease of implementation, where it allows researchers to leverage commonly used treatment effect estimation approaches. Such an approach would not possess the relative ease of implementation enjoyed by the proposed control variates estimator, which already demonstrates competitive efficiency relative to the oracle estimator in our simulations.

\subsection*{Robustness Conditions for $\hat \tau_\cv$}

\subsubsection*{Consistency}
When $\hat \tau_\cv$ is constructed so that $\hat \tau_\val$, $\hat \tau_\val^\ep$ are based on the efficient generalization estimators (\ref{eq:val-dr}) and (\ref{eq:ep-dr-val}), and $\hat \tau_\main^\ep$ is based on the AIPW estimator (\ref{eq:ep-dr}), $\hat \tau_\cv$ will inherit the robustness properties of its component estimators. Analogous to the proof of Theorem \ref{thm:complex-val}, we assume that all nuisance components are estimated from a separate held-out sample to simplify the asymptotic analysis of $\hat \tau_\cv$. Focusing on $\hat \tau_\val$, conditions developed in \cite{zeng2023efficient} imply that
$$
\hat \tau_\val - \tau_\tate = O_\PP \left( \frac{1}{\sqrt n} + ||\hat \mu_a - \mu_a|| \cdot ||\hat \pi -  \pi || + ||\hat \mu_a - \mu_a|| \cdot ||\hat \kappa - \kappa|| \right),
$$
with an analogous condition holding for $\hat \tau_\val^\ep$:
$$
\hat \tau_\val^\ep - \tau^\ep = O_\PP \left( \frac{1}{\sqrt n} + ||\hat \mu_a^\ep - \mu_a^\ep|| \cdot ||\hat \pi^\ep -  \pi^\ep || + ||\hat \mu_a^\ep - \mu_a^\ep|| \cdot ||\hat \kappa - \kappa|| \right).
$$
Similarly, as $\hat \tau_\main^\ep$ is an AIPW estimator, it is well-established that 
$$
\hat \tau_\main^\ep - \tau^\ep = O_\PP \left( \frac{1}{\sqrt n} + ||\hat m_a^\ep - m_a^\ep|| \cdot ||\hat g^\ep -  g^\ep || \right)
$$
Critically, these conditions imply that $\hat \tau_\val$ will be consistent so long as either
\begin{enumerate}
    \item $\hat \mu_a$ is correctly specified, or
    \item $\hat \mu_a$ is incorrectly specified, but $\hat \pi$ and $\hat \kappa$ are correctly specified,
\end{enumerate}
with an analogous condition holding for $\hat \tau_\val^\ep.$ Similarly, $\hat \tau_\main^\ep$ will be consistent so long as one of $m_a^\ep$ or $g^\ep$ are correctly specified. These collective conditions imply one only needs to get a subset of all nuisance models involved in constructing the control variates estimator correct.

\subsubsection*{Rate Robustness}
The bounds above also allow one to quantify the \textit{rates} at which all nuisance models need to be estimated in order to obtain parametric $\sqrt n$ rates of consistency and asymptotic normality. Particularly when there is insufficient subject matter knowledge to justify the choice of pre-defined parametric model classes for all nuisance functions, fitting all nuisance components with flexible machine learning methods can mitigate the risk of model misspecification.  Notice that $\hat \tau_\val$ will be $\sqrt n$ consistent and asymptotically normal if 
\begin{enumerate}
    \item $||\hat \mu_a - \mu_a|| \cdot ||\hat \pi -  \pi || = o_\PP(1/\sqrt{n})$
    \item $||\hat \mu_a - \mu_a|| \cdot ||\hat \kappa -  \kappa || = o_\PP(1/\sqrt{n})$
\end{enumerate}
Both conditions will hold if, for instance, $||\hat \mu_a - \mu_a|| = o_\PP(n^{-1/4})$, $||\hat \pi -  \pi || = o_\PP(n^{-1/4})$, and $||\hat \kappa -  \kappa || = o_\PP(n^{-1/4})$. Notice the above two conditions imply that within pairs, if one nuisance estimator is inconsistently estimated, then the other must attain parametric $\sqrt n$ rates of convergence to ensure asymptotic linearity. We note that $n^{-1/4}$ rates are attainable for a wide range of flexible machine learning algorithms, such as the highly adaptive LASSO developed in \cite{benkeser2016highly}.
\\ \\
An analogous condition holds for $\hat \tau_\val^\ep$, and through similar logic notice $\hat \tau_\main^\ep$ will be $\sqrt n$ consistent and asymptotically normal if both $||\hat m_a^\ep - m_a^\ep|| = o_\PP(n^{-1/4})$ and $||\hat g_a^\ep - g_a^\ep|| = o_\PP(n^{-1/4})$. We again emphasize such rates are attainable for a large class of flexible machine learning methods.

\subsection*{Proof of Theorem \ref{thm:complex-val}}
\subsubsection*{Asymptotic Linearity of $\hat \tau_\val^\text{IPSW}$ and $\hat \tau_\val^\text{IPSW,e.p.}$}
Without loss of generality, we will focus our attention on $\hat \tau_\val$. Analogous reasoning will establish all results for $\hat \tau_\val^\ep$. First, suppose the following regularity conditions hold:
\begin{enumerate}
    \item $|| \hat m_a - m_a || = o_\PP(n^{-1/4})$
    \item $|| \hat g - g || = o_\PP(n^{-1/4})$
    \item  $|| \hat m_a^\ep - m_a^\ep || = o_\PP(n^{-1/4})$
    \item $|| \hat g^\ep - g^\ep || = o_\PP(n^{-1/4})$
\end{enumerate}
To satisfy empirical process conditions we additionally assume that all nuisance models are fit in a separate held-out sample (\citealt{kennedy2020efficient}). One can implement cross-fitting methods to recover full efficiency of the resulting estimators.
\\ \\
We aim to show that $\hat \tau_\val^\text{IPSW}$ is asymptotically linear for $\tau_\tate$.
Recall that
$$
\hat 
\varphi^\text{IPSW}(Y,A,\bX,S) = \frac{S}{\kappa(\bX,A,Y,S)}
\left\{
\hat m_1(\bX) - \hat m_0(\bX) + \left(\frac{A}{\hat g(\bX)} - \frac{1-A}{1-\hat g(\bX)} \right)\hat m_{A}(\bX)
\right\}.
$$
Let
$$
\hat \phi(Y,A,\bX,S) = 
\hat m_1(\bX) - \hat m_0(\bX) + \left(\frac{A}{\hat g(\bX)} - \frac{1-A}{1-\hat g(\bX)} \right)\hat m_{A}(\bX),
$$
denote the uncentered efficient influence curve of the ATE functional $\psi = \E[\E(Y|A=1,\bX) - \E(Y|A=0,\bX)]$ under the full data structure. Then, the efficient influence curve under the observed data structure, denoted $\chi(\bm O)$, can be written as a function of the underlying full-data influence function (\citealt{rose2011targeted}):
\begin{align*}
\chi(\bm O) &=
\frac{S}{\kappa(\bX,A^*,Y)}\phi(Y,A,\bX,S) - \left(\frac{S}{\kappa(\bX,A^*,Y)} - 1 \right)\E[\phi(Y,A,\bX,S) | \bX,A^*,Y] - \psi \\
&= 
\varphi^\text{IPSW}(Y,A,A^*,\bX,S) - \left(\frac{S}{\kappa(\bX,A^*,Y)} - 1 \right)\E[\phi(Y,A,\bX,S) | \bX,A^*,Y] - \psi.
\end{align*}
Consider the estimator
$$
\hat \psi = \frac{1}{n}\sum_{i=1}^n \left(
\hat \varphi^\text{IPSW}(Y_i,A_i,A_i^*,\bX_i,S_i) - \left(\frac{S_i}{\kappa(\bX_i,A_i^*,Y_i)} - 1 \right)\hat\E[\phi(Y_i,A_i,\bX_i,S_i) | \bX_i,A_i^*,Y_i] \right),
$$
and further define 
\begin{equation}
\label{bigphi}
    \Phi(\bX,A^*,Y) = \E[ \phi(Y,A,\bX,S) | \bX,A^*,Y].
\end{equation}
Through Proposition 5 of \cite{levis2022double}, it has been shown that two-stage sampling estimators of this form have the following bias structure:
\begin{equation}
\label{dr-biased-samp}
\E[\hat \psi - \tau_\tate] = O_{\PP}\left(\frac{1}{\sqrt n} + ||\hat m - m || \cdot ||\hat g - g|| + ||\hat \kappa - \kappa || \cdot || \hat \Phi - \Phi||\right).
\end{equation}
Crucially, (\ref{dr-biased-samp}) implies that under the earlier regularity conditions 1 and 2, $\hat \psi$ is asymptotically linear with influence function $\chi(\bm O)$, since we assume $\kappa$ is known or can be estimated at a $\sqrt n$ rate.  
Similar to the finding in  \cite{rose2011targeted}, note that since the sampling probabilities are known in our setting, notice that for any estimated $\hat \Phi(Y,A,\bX,S)$, 
\begin{equation}
\label{mean-zero}
\E\left[\left(\frac{S}{\kappa(\bX,A^*,Y)} - 1 \right)\hat \Phi(\bX,A^*,Y) \right] = 0.
\end{equation}
Notice that $\hat \tau_\val^\text{IPSW}$ can be viewed as a special case of $\hat \psi$ which sets $\Phi(\bX,A^*,Y)=0$. (\ref{mean-zero}) then implies that $\hat \tau_\val = \frac{1}{n}\sum_{i=1}^n \hat \varphi^\text{IPSW}(Y_i,A_i,A_i^*,\bX_i,S_i)$ is asympotically linear for $\tau_\tate$ with influence function $ \varphi^\text{IPSW}(Y,A,A^*,\bX,S) - \tau_\tate$. 
\\ \\
With the conditions for asymptotic linearity of both estimators established, we briefly comment on the conditions under which each estimator is \textit{consistent} for $\tau_\tate$ and $\tau^\ep$.
Given the form in (\ref{dr-biased-samp}), notice that $\hat \tau_\val^\text{IPSW}$ will be consistent if either of $\hat m$ or $\hat g$ are correctly specified, which is the ``double-robustness" property expected of the AIPW estimator in standard non-missing data settings. An analogous property holds for $\hat \tau_\val^\text{IPSW,e.p.}$. 

\subsubsection*{Asymptotic Result}

We have demonstrated that  $\hat \tau_\val^\text{IPSW}$ and $\hat \tau_\val^\text{IPSW,e.p.}$ are both asymptotically linear with influence functions $\varphi^\text{IPSW}(Y,A,A^*,\bX,S) - \tau_\tate$ and $\phi^\text{IPSW,e.p.}(Y,A^*,\bX,S) - 0$, respectively. Under  regularity conditions 3 and 4, we also have that when $\hat \tau_\main^\ep$ is obtained as in (\ref{eq:ep-dr}), it is asymptotically linear with influence function $\phi^\ep(Y,A^*,\bX)-\tau^\ep$. By the asymptotic linearity of  $\hat \tau_\main^\ep$ and $\hat \tau_\val^\text{IPSW,e.p.}$, notice that
\begin{align*}
    \hat \tau_\val^\text{IPSW,e.p.} - \hat \tau_\main^\ep &= \frac{1}{n}\sum_{i=1}^n (\phi^\text{IPSW,e.p.}(Y_i,A_i^*,\bX_i,S_i) - \phi^\text{e.p.}(Y_i,A_i^*,\bX_i)) + o_\PP(1) \\
    &=
    \frac{1}{n}\sum_{i=1}^n \left(\frac{S_i}{\kappa(\bX_i,A_i^*,Y_i)}\phi^\text{e.p.}(Y_i,A_i^*,\bX_i) - \phi^\text{e.p.}(Y_i,A_i^*,\bX_i)\right) + o_\PP(1) \\
    &=
  \frac{1}{n}\sum_{i=1}^n \left(\frac{S_i}{\kappa(\bX_i,A_i^*,Y_i)} -1\right)\phi^\text{e.p.}(Y_i,A_i^*,\bX_i)+ o_\PP(1).  
\end{align*}
The joint asymptotic distribution of $\hat \tau_\val^\text{IPSW} - \tau_\tate$ and $\hat \tau_\val^\text{IPSW,e.p.} - \hat \tau_\main^\ep$ immediately follows by noting from the asymptotic linearity of $\hat \tau_\val^\text{IPSW}$,
$$
\hat \tau_\val^\text{IPSW} = \frac{1}{n}\sum_{i=1}^n (\varphi^\text{IPSW}(Y_i,A_i,A_i^*,\bX_i,S_i) - \tau_\tate) + o_\PP(1).
$$

\subsection*{Proof of (\ref{gen-functional})}

To establish (\ref{gen-functional}) under Assumptions \ref{sutva}-\ref{unconfoundedness} and \ref{source-pos}, we identify a generic counterfactual mean $\E[Y(a)]$, noting the proof follows by taking a contrast. We first note that Assumption \ref{eff-mods} which states $(Y,A) \indep S | \bX$ additionally implies $Y  \indep S | \bX, A$.
This holds since
\begin{align*}
    \PP(Y,S | A, \bX) &= \frac{\PP(Y,A,S|\bX)}{\PP(A|\bX)} \\
    &= \frac{\PP(Y,A|\bX)\PP(S|\bX)}{\PP(A|\bX)} \\
    &= \PP(Y|\bX,A)\PP(S|\bX) \\
    &= \PP(Y|\bX,A)\PP(S|\bX, A),
\end{align*}
where the final line establishes $Y \indep S | \bX, A$. Above, lines 2 and 4 hold by Assumption \ref{eff-mods}.
\\ \\ 
Now, notice 
\begin{align*}
    \E[Y(a)] &= \E[\E(Y(a) | \bX)] \\
    &= \E[\E(Y(a) | \bX, A=a)] \\
    &= \E[\E(Y | \bX, A=a)] \\
    &= \E[\E(Y | \bX, A=a, S=1)]. \\
\end{align*}
Above, line 2 above holds by Assumption 3, while line 4 holds by the corollary of Assumption \ref{eff-mods} provided above. 


\clearpage

\section*{Web Appendix B: Bootstrap Variance Estimation Details}

Here, we outline the general procedure for obtaining bootstrap estimates of $v$, $\Gamma$ and $V$. The procedure we propose is similar to the ones presented in \cite{guo2022multi} and \cite{yang2019combining}. Our procedure differs in that, rather than sampling with replacement from both the validation and main datasets, we only sample with replacement from the main dataset. The bootstrap procedure can be repeatedly applying the following two steps for $b \in \{ 1,\ldots,B\}$, where $B$ is the total number of iterations selected by the researcher:
\begin{enumerate}
    \item Sample $n$ subjects with replacement from the full sample, denoting the bootstrap sample by $\Scal^{(b)}$
    \item Given the bootstrap sample $\Scal^{(b)}$, obtain estimates $\hat \tau_\val^{(b)}$, $\hat \tau_\val^{\ep,(b)}$ and $\hat \tau_\main^{\ep,(b)}$ using the  same estimation procedure taken to obtain the original point estimates
\end{enumerate}

After repeating the above two steps $B$ total times, estimates can be obtained as
\begin{align*}
\hat v &= \frac{1}{B-1} \sum_{b=1}^B (\hat \tau_\val^{(b)} - \hat \tau_\val )^2 \\
\hat \Gamma &= \frac{1}{B-1} \sum_{b=1}^B (\hat \tau_\val^{(b)} - \hat \tau_\val)(\hat \tau_\val^{\ep,(b)} - \hat \tau_\main^{\ep,(b)} - (\hat \tau_\val^\ep - \hat \tau_\main^\ep)) \\
\hat V &= \frac{1}{B-1} \sum_{b=1}^B (\hat \tau_\val^{\ep,(b)} - \hat \tau_\main^{\ep,(b)} - (\hat \tau_\val^\ep - \hat \tau_\main^\ep))^2
\end{align*}
As demonstrated by \cite{guo2022multi}, so long as $\hat \tau_\val, \hat \tau_\val^\ep$ and $\hat \tau_\main^\ep$ are RAL, the bootstrap procedure yields consistent estimators of $\Gamma$ and $V$.

\clearpage

\section*{Web Appendix D: Simulation Details}
\label{simmy_details}

\begin{table}[h!]
    \centering
    \begin{tabular}{lll}
    \toprule
Parameter & Description & Value(s)  \\
\midrule 
$(\alpha_0, \bm \alpha)$  & Treatment model coefficients & $(0.1,-0.5,0.3,0.85)$ \\ 
$\zeta$ & Specificity & $0.95$ \\
$\tau$ & Baseline treatment effect & 1 \\
$(\beta_0, \bm \beta)$ & Outcome model coefficients & $(0,1,-3,0.5)$ \\
$\bm \gamma$ & Interaction effects & $(0.2, 0.4, -0.6)$ \\
$\varepsilon$ & Outcome cond. variance & 1 \\ 
$\bm \Sigma_{\bX}$ & Covariance matrix & $\begin{pmatrix}
    1 & 0.25  & 0.5 \\
   0.25 & 1 & -0.4 \\
   0.5 & -0.4 & 1
\end{pmatrix}$ \\ 
\midrule
$(\eta_0, \bm \eta)$ & Selection model coefficients & $(0,0.1,-0.2,0.6)$ or $\bm 0$  \\
$\rho$ & Relative size of validation data & $0.1,0.2,0.3,0.4,0.5$ \\
$\delta$ & Sensitivity & $0.95, 0.90, 0.85, 0.80$ \\
\bottomrule
    \end{tabular}
    \caption{Simulation parameter values. $\eta_0, \bm \eta$, $\rho$ and $\delta$ vary across simulation scenarios, other parameters remain fixed.}
    \label{sim-params}
\end{table}

\subsubsection*{Multiple Imputation}
The multiple imputation procedure is implemented using the \texttt{mice} package. To allow for robustness to outliers/flexibility in the imputation procedure, we use predictive mean matching (implemented with the \texttt{pmm} option). 

\subsubsection*{Treatment effect estimators}
For all five methods considered and implemented in the simulation study (including the construction of error-prone treatment effect estimators comprising the control variates) we use AIPW, implemented via the \texttt{AIPW} package, to estimate treatment effects. Nuisance functions are modeled via ensemble learning, implemented with the \texttt{SuperLearner} package and the following libraries: \texttt{SL.mean, SL.glm, SL.glm.interaction}. For more complex data-generating processes, we could extend this library to include more data-driven algorithms that make few if any assumptions about true nuisance function, such as \texttt{SL.ranger} and \texttt{SL.xgboost}, though we do not explore this here.

\subsection*{Control Variates Method}
In the variance reduction step, we estimate $\Gamma$ and $V$ using the empirical estimates of the asymptotic formulae presented in Theorem \ref{thmgenz}.

\clearpage

\section*{Web Appendix E: Data Application Details}

In this section, we detail the implementation of our data application with the VCCC database.

\subsection*{Generation of Semi-Synthetic Analysis Datasets}

We considered validation data sizes ranging from 10\% up to 50\%, in increments of 5\%. At each validation data size, we generated 1{,000} analysis datasets according to the process outlined below. For each fixed sampling probability $\rho \in \{0.1,\ldots,0.5\}$, we repeat the following 1{,}000 times
\begin{enumerate}
    \item We obtained a simple random sample of validated exposure measurements by generating a random variable $S \sim \text{Bernoulli}(\rho)$. Observations with $S=1$ are treated as validated, while for those with $S=0$ we proceed as if we do not have access to the true validated exposure measurements.
    \item To ensure variation in the outcome of interest, and allow for a benchmark to compare our estimator to, we generated synthetic 5-year survival outcomes $\tilde Y$ such that $\tilde Y | \bX, \bm C, A \sim \text{Bernoulli}(\text{expit}(\alpha A_i + A_i \bX_i^\top \bm \gamma + \bm C_i^\top \bm \beta))$. The model $\text{expit}(\alpha A_i + A_i \bX_i^\top \bm \gamma + \bm C_i^\top \bm \beta)$ is fit once before generating any of the 1{,}000 datasets by using the validated dataset. To address issues with censoring, we restrict the sample used to fit this initial model to only include subjects who initiated care at the VCCC up to and including 2006. Since the data were formed in 2011, any patients who first visit the VCCC later than 2006, but survive past 2011, would  have a censored 5-year survival outcome by construction. To make use of the otherwise complete data, we simulate synthetic outcomes for \textit{all patients}, including those with an initial visit later than 2006.
\end{enumerate}
For each $\rho$, this procedure yields us 1{,000} analysis datasets where (1) the set of validated exposure measurements we treat as available varies across the 1{,}000 datasets, (2) the synthetic outcome $\tilde Y$ varies across datasets, and (3) the covariates $(\bm C, \bX)$ are fixed across datasets.

We include the following variables in our data application

\begin{table}[h!]
    \centering
    \begin{tabular}{ll}
    \toprule
      Variable   & Component  \\
    \midrule
    5-year survival     & Outcome, $Y$ \\
    \midrule
    AIDS-defining event (ADE) at baseline & Exposure, $A$ \\
    \midrule
    Sex & Discrete covariate, $\bX$ \\
    Man who has sex with men (MSM) indicator & Discrete covariate, $\bX$ \\
    Injection drug use & Discrete covariate, $\bX$ \\
    Race & Discrete covariate, $\bX$ \\ 
    Ethnicity & Discrete covariate, $\bX$ \\ 
    Initiated ART within 1 month of first visit & Discrete covariate, $\bX$ \\
    \midrule
    Age at first visit (years) & Continuous covariate, $\bm C$ \\
    \bottomrule
    \end{tabular}
    \caption{VCCC data variables}
    \label{tab:vccc-variables}
\end{table}

\subsection*{Implementation}

For each $\rho$ and each of the 1{,}000 semi-synthetic analysis datasets, we implemented the control variates estimator with the method outline in Section \ref{obtaining-tauval}. Specifically, we obtained $\hat \tau_\val$ and $\hat \tau_\val^\ep$ with the generalization estimators outlined in Section \ref{obtaining-tauval}, with the only difference being that $\hat \tau_\val$ uses the validated exposure measurements and $\hat \tau_\val^\ep$ the error prone ones. $\hat \tau_\main^\ep$ is obtained through AIPW, using the error-prone exposures. All nuisance functions are estimated with a Super Learner, using the libraries \texttt{SL.mean}, \texttt{SL.glm} and \texttt{SL.glm.interaction}. $\Gamma$ and $V$ are estimated by using the influence functions of all component estimators, as outlined in Section \ref{forming-cv}.

\section*{Web Appendix F: Additional Simulation Results}

In this section we report the results of additional simulation exercises.

\subsection*{Altering the Overall Sample Size}

Continuing to consider the data-generating process in Section \ref{sec:simmy-setting}, we additionally vary the overall sample size $n$, considering values in the set $\{1000,2500,5000,10000\}$, recalling results for $n=5000$ are reported in the main text. We report the results in Figures \ref{fig:cv-srandom-1000}-\ref{fig:cv-scondrandom-10000}. Relative to the main text, where we set $n=5000$, the results are qualitatively similar across all 3 additional sample sizes.

\foreach \n in {1000,2500,10000} {
  \begin{figure}[h!]
    \centering
    \includegraphics[width=0.8\textwidth]{paper_figures/cv-srandom\n.pdf}
    \caption{Simulation results with $n=\n$ and validation data obtained completely at random.}
    \label{fig:cv-srandom-\n}
  \end{figure}
}

\foreach \n in {1000,2500,10000} {
  \begin{figure}[h!]
    \centering
    \includegraphics[width=0.8\textwidth]{paper_figures/cv-scondrandom\n.pdf}
    \caption{Simulation results with $n=\n$ and validation data obtained conditionally at random.}
    \label{fig:cv-scondrandom-\n}
  \end{figure}
}

\clearpage

\subsection*{Mis-specification of Nuisance Functions}

\begin{figure}[h!]
    \centering
    \includegraphics[scale=0.7]{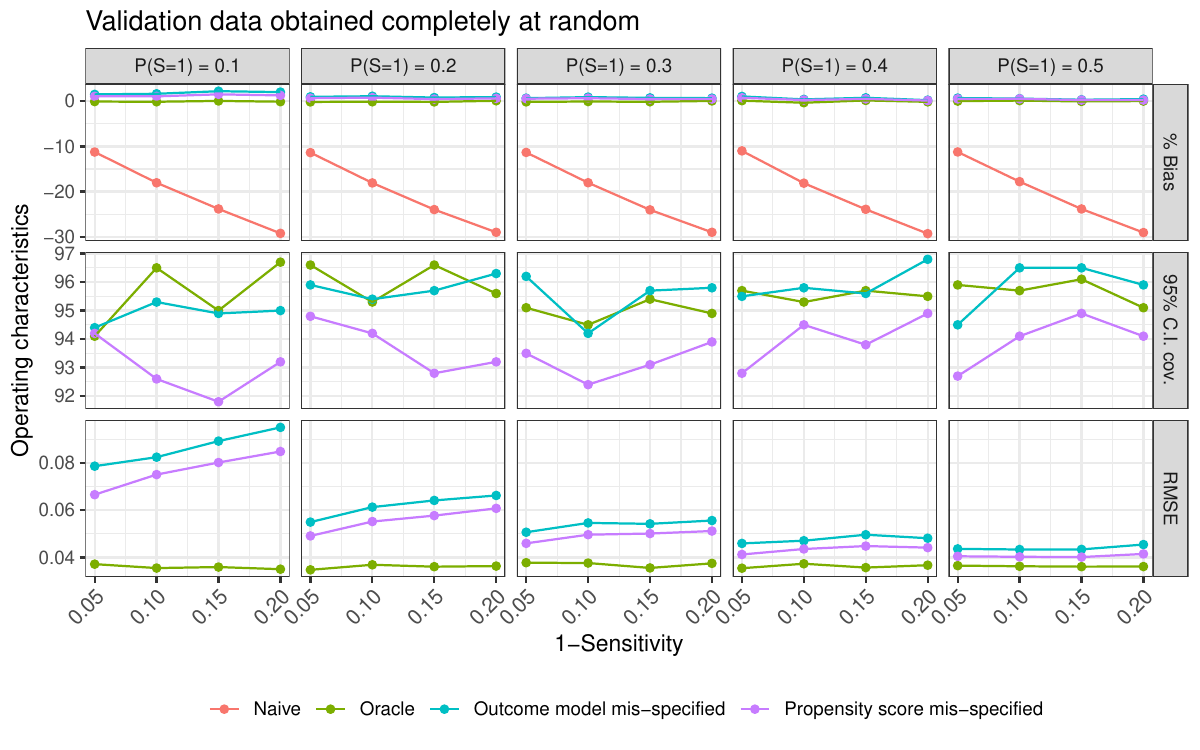}
    \caption{Simulation results for $\hat \tau_\cv$ when one of $\hat \mu_a$ and $\hat \pi$ is mis-specified, and validation data is obtained completely at random. The overall sample size $n$ is fixed at 5{,}000.}
    \label{fig:dr-srandom}
\end{figure}

\begin{figure}[h!]
    \centering
    \includegraphics[scale=0.7]{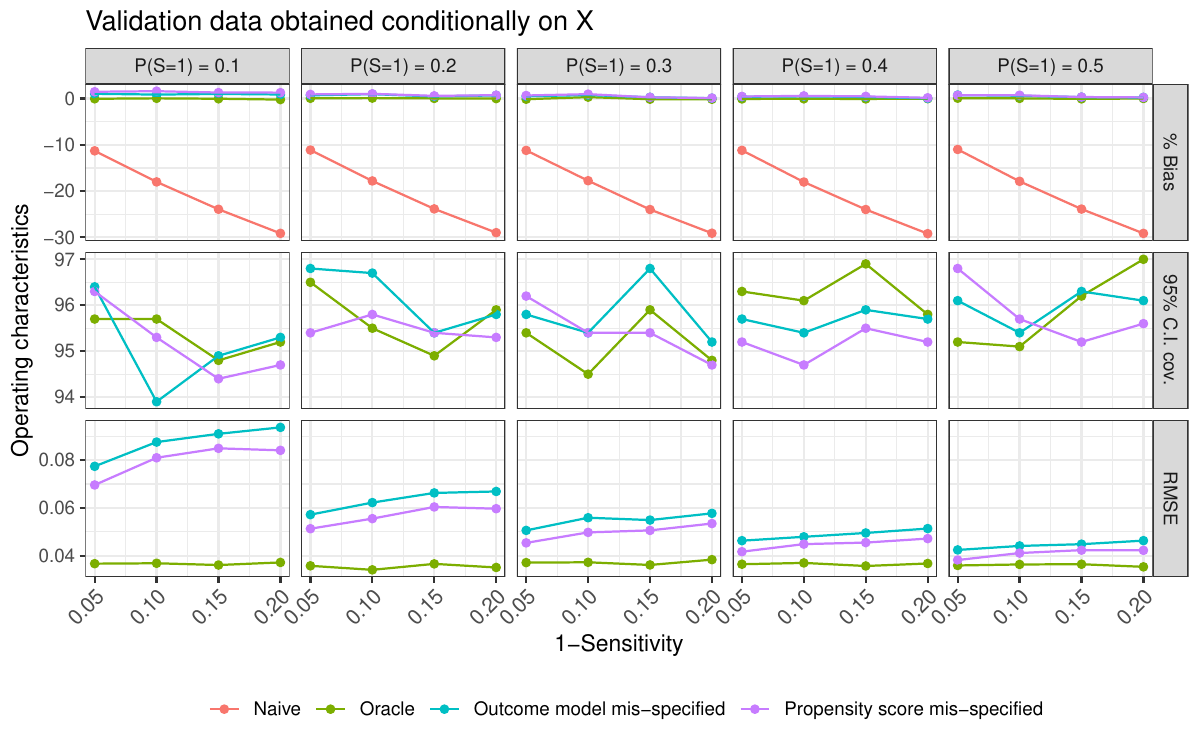}
    \caption{Simulation results for $\hat \tau_\cv$ when one of $\hat \mu_a$ and $\hat \pi$ is mis-specified, and validation data is obtained completely at random. The overall sample size $n$ is fixed at 5{,}000.}
    \label{fig:dr-scondrandom}
\end{figure}

To exhibit the double-robustness properties of the control variates estimator, we perform simulations in which we intentionally mis-specify nuisance functions. Specifically, we continue to consider the data-generating process outlined in \ref{sec:simmy-setting}, and investigate the performance of the control variates estimator when
\begin{itemize}
    \item The outcome regression model $\hat \mu_a(\bX)$ is mis-specified, or
    \item The propensity score $\hat \pi(\bX)$ is mis-specified 
\end{itemize}
We enforce mis-specification of the outcome regression model by fitting a linear model that omits interactions (through the SuperLearner library \texttt{SL.glm}), and mis-specification of the propensity score model by using the SuperLearner library \texttt{SL.mean}. Figures \ref{fig:dr-srandom}-\ref{fig:dr-scondrandom} plot the results of this exercise. As the primary focus of this exercise is to determine whether the proposed method is robust to model mis-specification, we compare these two versions of a partially mis-specified $\hat \tau_\cv$ to the oracle and naive estimators discussed in Section \ref{simmy}. We see that in both cases, $\hat \tau_\cv$ remains consistent. In terms of efficiency, mis-specification of the outcome regression $\mu_a(\bX)$ causes greater deterioration in efficiency, relative to mis-specification of $\hat \pi(\bX)$.


\subsection*{Complex Validation Data Sampling Schemes}

In this section, we study scenarios presented in Section \ref{sec:complex-val}, and their associated estimators. Specifically, we consider a data-generating process analogous to the one presented in Section \ref{sec:simmy-setting}:
\begin{align*}
\bX_i &\sim N(\bm 1, \bm \Sigma_{\bX})
& (\text{Covariates}); \\
A_i | \bX_i &\sim \text{Bernoulli}(\pi(\bX_i)), \enskip \pi(\bX_i) = \text{expit}(\alpha_0 + \bX_i^\top \bm \alpha) & (\text{Exposure}); \\
A_i^* | A_i &\sim \text{Bernoulli}(p_i),  \ p_i = A_i \delta + (1-A_i)\zeta & (\text{Measurement}); \\
S_i | \bX_i &\sim \text{Bernoulli}(\kappa(\bX_i)), \enskip \kappa(\bX_i) = \frac{\rho \cdot \text{expit}(\eta_0 + \bm Z_i^\top \bm \zeta)}{\frac{1}{n}\sum_{k=1}^{n} \text{expit}(\eta_0 + \bm Z_k^\top \bm \zeta)} & (\text{Val. data selection}); \\
Y_i | A_i, \bX_i &\sim N(\mu(\bX_i), \varepsilon), \enskip \mu(\bX_i) = \beta_0 + \tau A_i + \bX_i^\top \bm \beta + A_i\bX_i^\top \bm \gamma & (\text{Outcome}),
\end{align*}
where $\bm Z = (\bX, Y, A^*)$. The key distinction with the setting considered in Section \ref{sec:simmy-setting} is the validation data sampling mechanism, which here depends on $\bm Z$ rather than only $\bX$. We set $\bm \zeta = (\bm \eta, 0.25,0.25)$, where $\bm \eta$ are the selection coefficients defined in Table \ref{sim-params}. Effectively, this choice of $\bm \zeta$ generates validation samples that over-sample observations with larger outcomes and error-prone exposure measurements. Similar to our main exercise, we set the total sample size $n=5{,}000.$
We employ the method proposed in Section \ref{sec:complex-val}, comparing it to the same oracle, naive, and validation-data only estimators considered in Section \ref{sec:simmy-setting}. 
\begin{figure}[h!]
    \centering
    \includegraphics[scale=0.6]{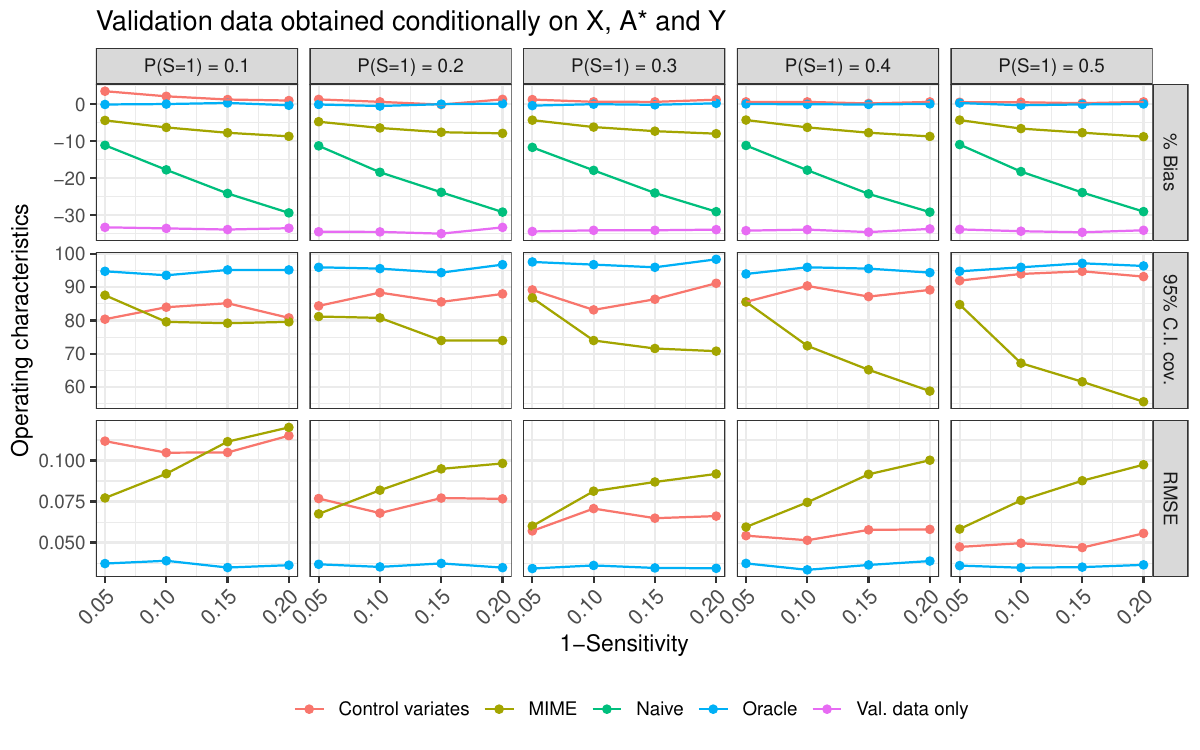}
    \caption{Simulation results for complex validation data sampling schemes. }
    \label{fig:complex-val-sim-res}
\end{figure}
\\ \\
The results are reported in Figure \ref{fig:complex-val-sim-res}. Similar to the simulations considered in Section \ref{simmy}, we see that multiple imputation is slightly biased due to a mis-specification of the imputation model, hampering its coverage and RMSE. Notably, outside of settings with relatively little measurement error and validation data, the control variates estimator tends to outperform multiple imputation in terms of RMSE. Similar to our main simulation exercises, a slight mis-specification in the predictive mean matching imputation model leads MIME to exhibit a small degree of bias.

\clearpage


\begin{landscape}
\thispagestyle{empty}
\section*{Web Appendix G: Additional Information}
\begingroup

\setlength{\tabcolsep}{10pt} 
\renewcommand{\arraystretch}{1.5} 
\newgeometry{,vmargin=.2cm,hmargin=.2cm} 

\begin{table}[]
    \centering
    \resizebox{\textwidth}{!}{
    \begin{tabular}{llll}
    \toprule
     Paper    & Relationship of interest & Measurement error variable & Validation data source  \\
     \midrule
     \cite{josey2023estimating}    & PM2.5 and adverse health outcomes & Grid-level PM2.5 concentrations & PM2.5 measurements in grids containing ground monitors \\
     \cite{spiegelman1997regression} & Breast cancer incidence and vitamin A intake & Self-reported vitamin A intake & Subsample of respondents whose responses were validated \\
     \cite{braun2017propensity} & Resection vs biopsy on brain cancer survival & Resection indicator & SEER-Medicare validation data \\
     \cite{lyles2011validation} & Bacterial vaginosis and various risk factors & Clinical bacterial vaginosis diagnoses & Lab-based diagnoses (considered gold-standard) \\
     \cite{lyles2007combining} & SIDS and maternal antibiotic use during pregnancy & Self-reported maternal antibiotic use & Medical records \\
     \cite{magaret2008incorporating} & HIV infection risk and various covariates & HIV acquisition (single screening)  &  Patients with multiple screening tests conducted \\
     \cite{lim2015measurement} & Physical activity and various outcomes & Reported physical activity & Subsample of patients provided accelerometers \\
     \cite{shepherd2023multiwave} & Maternal weight gain and childhood obesity & All variables in EHR database & Manual chart review \\
     \cite{amorim2024three} & Interaction between contraception and ART on pregnancies & ART regimen, contraceptions, pregnancy & Chart review and phone interviews \\
         \bottomrule
    \end{tabular}
    }
    \caption{Example set of papers making use of validation data sources to correct for measurement error in a main sample.}
    \label{valdataexamples}
\end{table}

\endgroup 
\end{landscape}
\restoregeometry



\end{document}